\documentclass[useAMS,usenatbib,twocolumn,fleqn]{mnras}
\usepackage[T1]{fontenc}
\usepackage{ae,aecompl}
\usepackage[export]{adjustbox}
\usepackage{natbib, graphics, epsfig}
\usepackage{stmaryrd}

\usepackage{times}
\usepackage{amsmath}
\usepackage{amssymb}
\usepackage{textcomp}
\usepackage{verbatim}

\newcommand{\msun}{{\,\rm M_\odot}}

\newcommand{\kms}{\,{\rm km}\,{\rm s}^{-1}}

\newcommand{\pc}{\,{\rm pc}}
\newcommand{\kpc}{\,{\rm kpc}}
\newcommand{\Mpc}{\,{\rm Mpc}}

\newcommand{\mstar}{M_{\star}}
\def\gsim{ \lower .75ex \hbox{$\sim$} \llap{\raise .27ex \hbox{$>$}} }
\def\lsim{ \lower .75ex \hbox{$\sim$} \llap{\raise .27ex \hbox{$<$}} }

 \title[Predictions for the high-redshift Universe in ETHOS]{ETHOS -- an effective theory of structure formation: predictions for the high-redshift Universe -- abundance of galaxies and reionization}
\author[M.~R.~Lovell et al.]{Mark R. Lovell\thanks{email: lovell@hi.is}$^{1,2,3}$, Jes\'us Zavala$^{1}$, Mark Vogelsberger$^{4}$\thanks{Alfred P. Sloan Fellow}, Xuejian Shen$^4$, \newauthor Francis-Yan Cyr-Racine$^{5}$, Christoph Pfrommer$^{6}$, Kris Sigurdson$^{7}$, \newauthor Michael Boylan-Kolchin$^{8}$, and Annalisa Pillepich$^{3}$\\
$^{1}$Center for Astrophysics and Cosmology, Science Institute, University of Iceland, Dunhagi 5, 107 Reykjavik, Iceland \\
$^{2}$Institute for Computational Cosmology, Durham University, South Road, Durham DH1 3LE, UK\\
$^{3}$Max-Planck-Institut f\"ur Astronomie, K\"onigstuhl 17, D-69117 Heidelberg, Germany \\
$^{4}$Department of Physics, Kavli Institute for Astrophysics and Space Research, Massachusetts Institute of Technology, Cambridge, MA 02139, USA\\
$^{5}$Department of Physics, Harvard University, Cambridge, MA 02138, USA\\
$^{6}$Heidelberg Institute for Theoretical Studies, Schloss-Wolfsbrunnenweg 35, 69118 Heidelberg, Germany\\
$^{7}$Department of Physics and Astronomy, University of British Columbia, Vancouver, British Columbia V6T 1Z1, Canada \\
$^{8}$Department of Astronomy, The University of Texas at Austin, 2515 Speedway, Stop C1400, Austin, TX 78712-1205, USA}

\date{Accepted *** Received ***; in original
  form ***} 
\pubyear{2017}

\begin{document}

\label{firstpage}
\pagerange{\pageref{firstpage}--\pageref{lastpage}} 
  
\maketitle
 
\begin{abstract}
We contrast predictions for the high-redshift galaxy population and reionization history between cold dark matter (CDM) and an alternative self-interacting dark matter model based on the recently developed ETHOS framework that alleviates the small-scale CDM challenges within the Local Group. We perform the highest resolution hydrodynamical cosmological simulations (a 36~Mpc$^3$ volume with gas cell mass of $\sim10^5\msun$ and minimum gas softening of $\sim180$~pc) within ETHOS to date -- plus a CDM counterpart -- to quantify the abundance of galaxies at high redshift and their impact on reionization. We find that ETHOS predicts galaxies with higher ultraviolet (UV) luminosities than their CDM counterparts and a faster build-up of the faint end of the UV luminosity function. These effects, however, make the optical depth to reionization less sensitive to the power spectrum cut-off: the ETHOS model differs from the CDM $\tau$ value by only 10 per cent and is consistent with Planck limits if the effective escape fraction of UV photons is 0.1-0.5. We conclude that current observations of high-redshift luminosity functions cannot differentiate between ETHOS and CDM models, but deep JWST surveys of strongly-lensed, inherently faint galaxies have the potential to test non-CDM models that offer attractive solutions to CDM's Local Group problems.

\end{abstract}


\section{Introduction}
\label{intro}
Evidence from the dynamics of gas and stars within galaxies, the large-scale
distribution of galaxies and baryonic matter, and the cosmic microwave
background has firmly established a paradigm in which gravitational forces in
the Universe are dominated by a component that has relatively low thermal
velocities at early times \citep[e.g.][]{White83,Viel13}. In addition to being
``cold'', this dark matter is generally assumed to be collisionless; this is a
central pillar of the dark energy plus cold dark matter ($\Lambda$CDM) model
that has seen many successes on cosmologically large scales
\citep[e.g.][]{Springel05b,Planck16} and is the basis of our current theory for
galaxy formation (\citealt{White78, blumenthal1984}; see
\citealt{somerville2015} and \citealt{naab2017} for recent reviews). Our ability to simulate the formation and evolution of structure -- including galaxies -- in
this model has increased dramatically, and
state-of-the-art simulations are now able to reproduce many properties of the
baryonic and total matter distribution over a variety of epochs in
cosmologically-representative volumes
\citep[e.g.,][]{Vogelsberger14,Genel14,Dubois14, Crain15,Schaye15}.

However, it is important to note that dark matter has only been detected by its
gravitational influence, meaning we only have upper limits on its primordial velocity dispersion and
collisionality via its effects on baryonic structures and the clustering of
matter. Given current constraints, it is certainly possible that dark matter is
neither fully cold nor collisionless. An essential question, therefore, is
whether dark matter deviates from the phenomenology of a cold and collisionless
particle on any scale relevant for astrophysical observations at any
cosmological epoch. While many of the most frequently-considered particle
candidates for dark matter are indeed cold and collisionless, including
weakly-interacting massive particles (WIMPs) and QCD axions \citep{feng2010},
there are diverse particle physics models in which dark matter has negligible
interactions with baryons while having a free-streaming length of $\sim
10\,$kpc, a self-scattering cross section that is comparable to the strength
characteristic of the strong nuclear force in the Standard Model of particle
physics ($\sim 10\,{\rm cm^2\,g^{-1}}$), or a combination of both.

Such models are also of interest astrophysically, in the context of attempts to
understand observations at sub-galactic scales. In this regime, agreement
between predictions from CDM models and observations is not established yet as it is on
large scales. Even in systems where dark matter dominates the gravitational
potential, astrophysical systems are generically less dense and less abundant
than naive CDM predictions (see \citealt{bullock2017} for a recent review).
These potential ``small-scale challenges'' -- the cusp-core, missing
satellite/field-dwarf, and too-big-to-fail problems \citep{Flores94, Moore94,
Moore99, Klypin99, Zavala09,Papastergis11,Klypin15,BoylanKolchin11} -- have
been the astrophysical motivation for considering dark matter models that
abandon the cold or collisionless assumptions of standard CDM.

These challenges might however be solved by the complex and not yet fully
understood physics of galaxy formation and evolution. In fact, simulations
rooted firmly in the CDM paradigm have demonstrated that star formation
feedback may be able to alleviate many of the apparent tensions facing the CDM
model. In particular, feedback-driven gravitational potential fluctuations can
heat dark matter \citep{Pontzen_Governato_11}, reducing central densities and
often forming dark matter cores \citep{Governato10,Zolotov2012, teyssier2013,
munshi2013, di-cintio2014, Onorbe15, chan2015, tollet2016, fitts2017,
read2017}. Thus, signatures of non-standard dark matter can be difficult to
disentangle from those of baryonic feedback. One possibility in the
low-redshift Universe is to study the internal kinematics of the least luminous galaxies possible ($\mstar
\la 10^6\,\msun$), as recent results indicate that feedback-induced cores will
be minimal or non-existent in such systems (\citealt{di-cintio2014, tollet2016,
fitts2017}, though see \citealt{Sawala16a} and \citealt{read2017}). Firmly
establishing definitive tests in the local and distant Universe are therefore of
crucial importance for understanding whether or not changes are required to the
CDM paradigm. 

Observationally, the power spectrum of dark matter is required to extend at
least down to the mass scale of dwarf galaxies ($M_{\rm dm} \approx
10^{10}\,\msun$) without exhibiting a damping signature, with constraints
coming from counts of satellite galaxies in the nearby Universe
\citep{Polisensky11,Lovell14,Kennedy14,Kim17} and structure in the Lyman-$\alpha$
forest at higher redshifts \citep[e.g.][]{Viel13,Irsic17}. Quoted limits at
high redshifts are typically sensitive to a number of effects (e.g.,
uncertainties in the thermal history of the Universe and the production
mechanism of dark matter; see, e.g., \citealt{Puchwein12, Bozek16, Garzilli17,murgia2017}), but it remains possible that the particle nature of dark matter
is important (or dominant) in setting the minimum scale for galaxy formation
through damping of primordial perturbations.

Dark matter self-interactions \citep[e.g.][]{Spergel00,FengJ09,Loeb11} affect
structure in a different manner: as opposed to suppressing structure in the
linear regime, dark matter self-interactions operate in the highly non-linear
regime of structure formation, affecting primarily the dense centres of dark
matter haloes. Such models have been explored in the context of structure
formation simulations for nearly two decades, under the umbrella term of
self-interacting dark matter \citep[SIDM; e.g.][]{Yoshida00,Dave01,Colin02}.
Recent years have seen a new generation of simulations demonstrating SIDM's
viability over the full range of scales relevant for galaxy formation and its
ability to mitigate outstanding CDM challenges at the scale of dwarf galaxies
via dark matter physics \citep[e.g.][]{Vogelsberger12,
Rocha13,Zavala13,Vogelsberger13b,Vegetti14, Vogelsberger14c,Elbert15,Dooley16, Kaplinghat16,Creasey17,Kamada17,
robles2017, Brinckmann18}. For a recent review on SIDM see \citet{Tulin18}.

The ETHOS framework~\citep{CyrRacine16,Vogelsberger16} generalises structure
formation theory to allow for non-gravitational interactions -- both
self-collisions (as in SIDM) and a cut-off in the primordial power spectrum
caused by hidden interactions between dark matter and relativistic particles in the
early Universe
\citep[e.g.][]{Hofmann2001,ChenKam2001,Boehm02,Green2004,Bertschinger2006,Bringmann2007h,vandenAarssen2012,CyrRacine:2012fz}. Such cut-offs are different in nature than the
free-streaming cut-off in Warm Dark Matter (WDM), but they also result in a
suppression of the abundance of galaxies, a helpful feature to explain the
observed dearth of dwarf galaxies (see \citealp{Boehm14} and
\citealp{Buckley14} for analyses of these models with simulations). In
\citet{Vogelsberger16}, using dark-matter-only simulations, a specific benchmark model (ETHOS-4) was found that is
consistent with current constraints from the population of Milky Way satellite
haloes and is able to reduce both the predicted abundance of satellites and
their dark matter densities relative to the predictions from CDM, thereby
addressing outstanding small-scale issues of the $\Lambda$CDM paradigm in a
compelling way.
   
This benchmark model, which we call ETHOS from this point onwards for simplicity, has been
calibrated to match broadly the observed satellite subhalo properties of the
Milky Way. Therefore, it has only been explored in the local Universe. In order
to fully assess the viability of this model we must however compare its
predictions to observables in other environments. One such regime is the
high-redshift Universe. The properties of galaxies at high redshift are
affected by the collapse time of their host haloes. This process is delayed in
dark matter models with primordial power spectrum cut-offs \citep[e.g. for
WDM,][]{Colin00,Bode01,Lovell12,Lovell16}.  In the ETHOS model, we therefore
expect an impact on the high-redshift mass and luminosity functions with
potentially detectable differences with respect to the CDM case. Furthermore, a
delayed collapse time, coupled to the lower number density of small galaxies,
should lead, at least naively, to lower star formation rates, which in turn result in a lower production
rate of high-energy, ionizing photons and subsequently to a later epoch of
reionization. If the delay in reionization is in severe tension with cosmic
microwave background (CMB) estimates \citep{PlanckXLVI16}, then the model is
ruled out. In the context of WDM, recent studies using semianalytic modelling
of the galaxy population have shown that this delay does occur, with the end
results being sensitive to assumptions about galaxy formation physics.
Particularly for reionization constraints, the role of strong high-redshift
starbursts in WDM leads to galaxies that are brighter in the UV than is the
case in CDM, therefore partially compensating for the deficit in the number of
galaxies \citep{Bose16c,Rudakovskyi16,Dayal17}.
   
In this paper, we confront the benchmark ETHOS model with constraints from the high-redshift Universe. We perform the first high resolution, cosmological, hydrodynamical simulations within the ETHOS framework, taking into account the matter power spectrum cut-off and the self-interactions of dark matter particles, while 
baryonic physics is incorporated in a state-of-the-art galaxy formation and evolution model. We obtain predictions for the high-redshift luminosity functions and the number density of ionizing photons. We compare these to a CDM simulation with the same initial conditions, phases and treatment of baryonic physics. We then assess our results in the context of current observational constraints to test the viability of the ETHOS model,  and present luminosity function predictions for the James Webb Space Telescope (JWST).

This paper is organised as follows. In Section~\ref{sec:sims} we describe our
methods for simulating the high-redshift galaxy population with CDM and ETHOS, while in Section~\ref{res} we present our results on the luminosity functions and optical depth for reionization. We present our conclusions in Section~\ref{conc}.

\section{Simulations}
\label{sec:sims}
   
We perform cosmological hydrodynamical simulations within CDM and ETHOS using
the {\sc Arepo} code \citep{Springel10} combined with a well-tested galaxy
formation model~\citep[][]{Vogelsberger13, Torrey14,
Vogelsberger14,Vogelsberger14b,Genel14, Weinberger17, Pillepich17}. The {\sc Arepo} code has been significantly extended to include isotropic and elastic
self-interactions \citep{Vogelsberger16} following arbitrary velocity-dependent
interaction cross sections. The ETHOS simulation employs the primordial power spectrum cut-off and self-interaction
cross section of the ETHOS-4 model presented in~\cite{CyrRacine16} and
\cite{Vogelsberger16}\footnote{See Table 1 in \citet{Vogelsberger16} for the particle physics and effective parameters of the ETHOS-4 model. Fig. 1 on that paper shows the linear power spectrum and the velocity-dependent transfer cross section for this model.}. The cosmological parameters for the simulations are:
matter density $\Omega_{0}=0.302$,  dark energy density
$\Omega_{\Lambda}=0.698$, baryon density $\Omega_{\rm b}=0.046$, Hubble
parameter $H_0 = 100\,h\,\kms \Mpc^{-1} = 69.1\,\kms \Mpc^{-1}$,  power
spectrum normalisation $\sigma_{8}=0.838$, and power spectrum slope index
$n_\rmn{s}=0.967$, which are consistent with recent Planck data
\citep{Planck14,Spergel15}. We perform simulations over a $(36.2\Mpc)^3$ periodic volume with a dark matter particle mass of $1.76\times10^{6}\msun$ and
a comoving dark matter softening length of $724\pc$. The average gas cell mass
is $2.69 \times 10^5\msun$ and the gas softening length is adaptive with a
comoving minimum of $181\pc$. These simulations are currently the best resolved uniform box hydrodynamical simulations of alternative dark matter models. 
 
\begin{figure}
\includegraphics[width=0.48\textwidth]{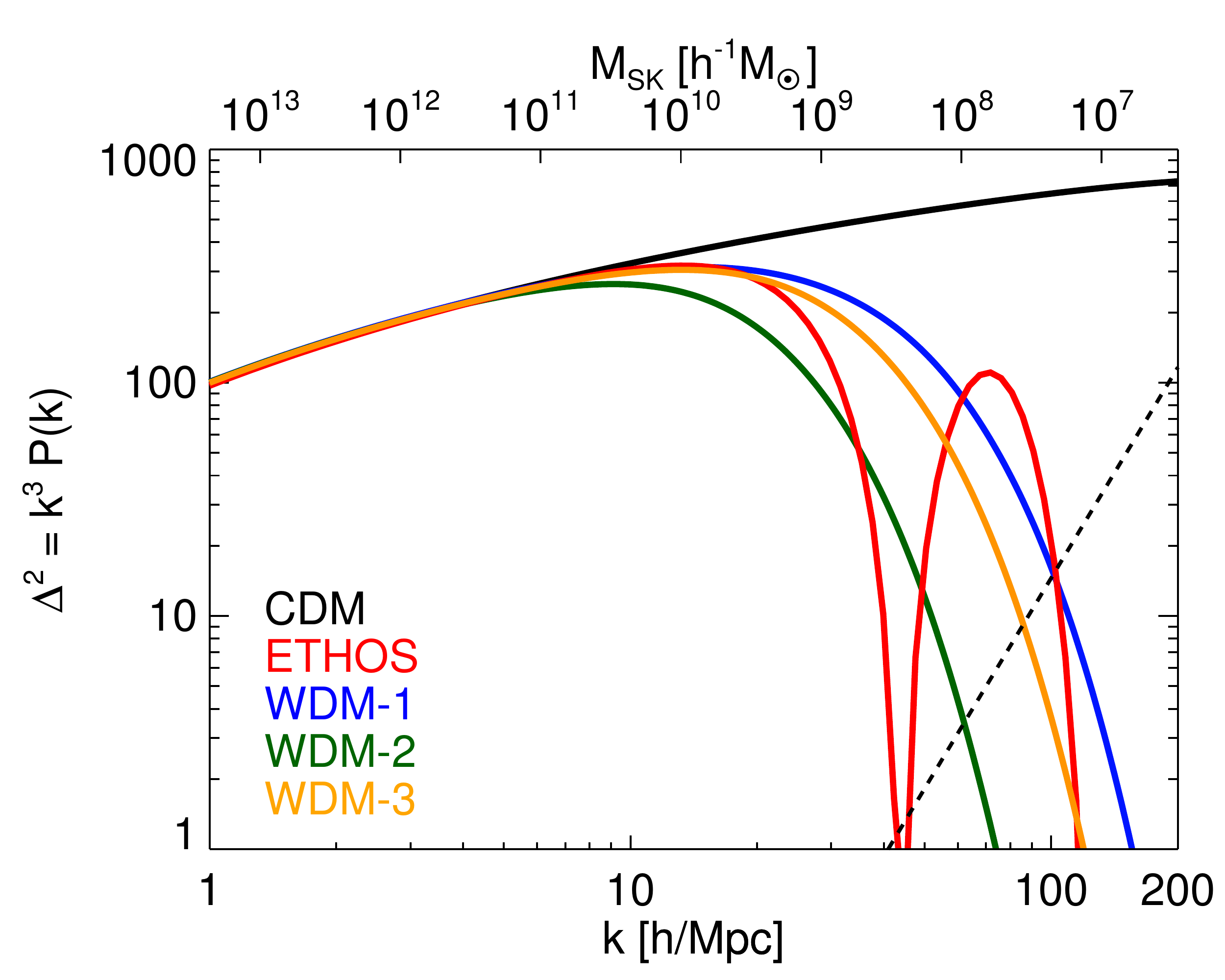}
\caption{Dimensionless linear matter power spectra for CDM and ETHOS, shown as
black and red solid lines, respectively. Also included are three warm dark matter models: WDM-1, WDM-2 and WDM-3. These are three sterile neutrino models in which the sterile neutrino mass is $7\,{\rm keV}$ and the lepton asymmetry is $L_6=8$ (blue), $L_6=11.2$ (green), and $L_6=8.9$ (orange) for WDM-1, WDM-2 and WDM-3 respectively. In the upper $x$-axis we show the halo mass scale associated with each wavenumber, using a sharp $k$-space cut-off. The dashed line shows the power spectrum due to Poisson noise in our simulations.}
\label{MPS}
\end{figure}    

The initial conditions of both simulations, CDM and ETHOS, share the same
random field, but the ETHOS initial conditions have a fluctuation spectrum with
an amplitude rescaled by the ETHOS linear matter power spectrum~\citep[taken
from][]{CyrRacine16}. This contains a small-scale primordial cut-off in the
power spectrum including dark acoustic oscillations (DAOs) due to the
interaction of the dark matter particles with the relativistic species (dark
radiation). This cut-off occurs at a wavenumber of 13~$h^{-1}\rmn{Mpc}$, which is remarkably similar to the cut-off wavenumber of the 7~keV sterile neutrino that could be responsible for the unidentified $3.5\,{\rm keV}$ emission line in some galaxy clusters and the Andromeda galaxy
\citep{Bulbul14,Boyarsky14a}. We present the ETHOS power spectrum in Fig.~\ref{MPS}, along with power spectra for three 7~keV sterile neutrino models with different free-streaming lengths, as originally presented in \citet{Lovell16,Lovell17b}. These models are characterized by the lepton asymmetry $L_6$, defined as $10^6$ times the difference in lepton and
anti-lepton abundance normalised by the entropy density. The $L_6=8$ model has
the shortest free-streaming length of any $7\,{\rm keV}$ sterile neutrino. On
the other hand, $L_6=11.2$ has the longest free-streaming length expected from
a $7\,{\rm keV}$ sterile neutrino responsible for the unidentified $3.5\,{\rm
keV}$ emission line in some galaxy clusters and the Andromeda galaxy
\citep{Bulbul14,Boyarsky14a}. The case with $L_6=8.9$ is in between and peaks
at the same wavenumber as the ETHOS benchmark model.  Qualitatively, the ETHOS model is similar to the cooler $7\,{\rm keV}$ sterile neutrino models,
including those that match the $3.5\,{\rm keV}$ line; the main difference being
the DAOs in ETHOS, which are not present in WDM models. The detailed parameter
calibration leading to the ETHOS benchmark model is described in
\citet{Vogelsberger16}.   
   
A well known problem for N-body simulations of models with a resolved cut-off
scale in the linear matter power spectrum is the spurious fragmentation of
filaments,  which is ultimately caused by the power spectrum due to shot noise in a simulation exceeding the small-scale power at wavenumbers higher than the physical cut-off of the model (see dashed line in Fig.~\ref{MPS}). The spurious fragments coalesce into haloes that could potentially host
galaxies, thus affecting our results. The characteristic mass scale below which
these spurious haloes form and dominate the mass function, $M_\rmn{lim}$, has
been shown by~\citet{Wang07} to be well described by $M_\rmn{lim} = 10.1 \,
\bar\rho \, d \, k_\rmn{peak}^{-2}$, where $\bar\rho$ is the mean density of
the Universe, $k_\rmn{peak}$ is the wavenumber at which the dimensionless
matter power spectrum attains its maximum amplitude, and $d$ is the mean
inter-particle separation of the simulation. For the purpose of spurious
fragmentation, ETHOS is similar to WDM with an equivalent peak in the
dimensionless power spectrum, as the amplitude of first dark acoustic peak is well below that of the main power spectrum peak. For the benchmark model we
analyse, we find: $M_\rmn{lim}\sim1.2\times10^8\msun$, or $\sim70$ particles
for the resolution of our simulations. We comment below on whether these
objects may have any effect on our results. 

\begin{figure*}
\setbox1=\hbox{\includegraphics[width=0.99\textwidth]{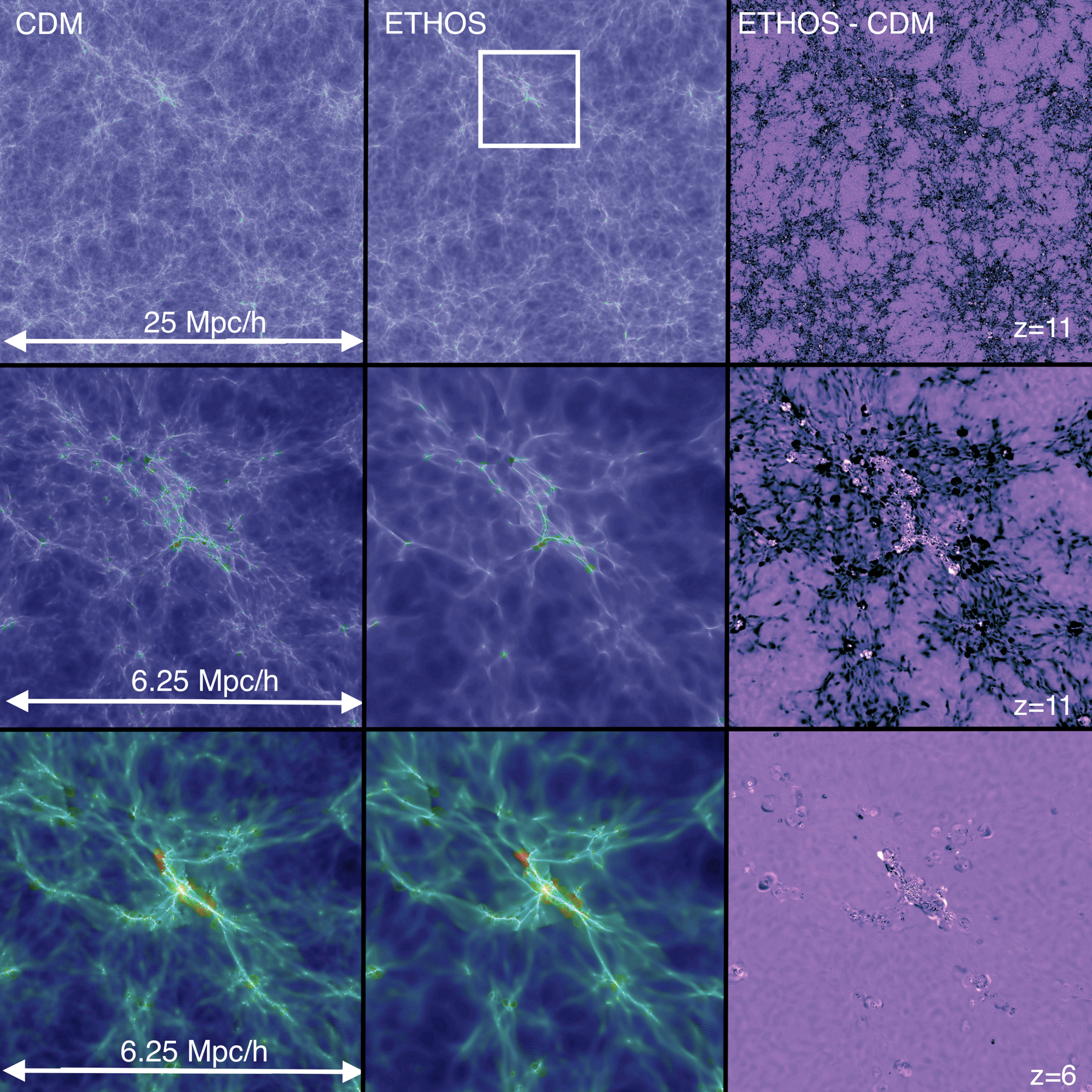}}
\includegraphics[width=0.99\textwidth]{NineFramesH3_Deg.png}\llap{\makebox[\wd1][r]{\raisebox{0.245\wd1}{\includegraphics[width=0.23\textwidth]{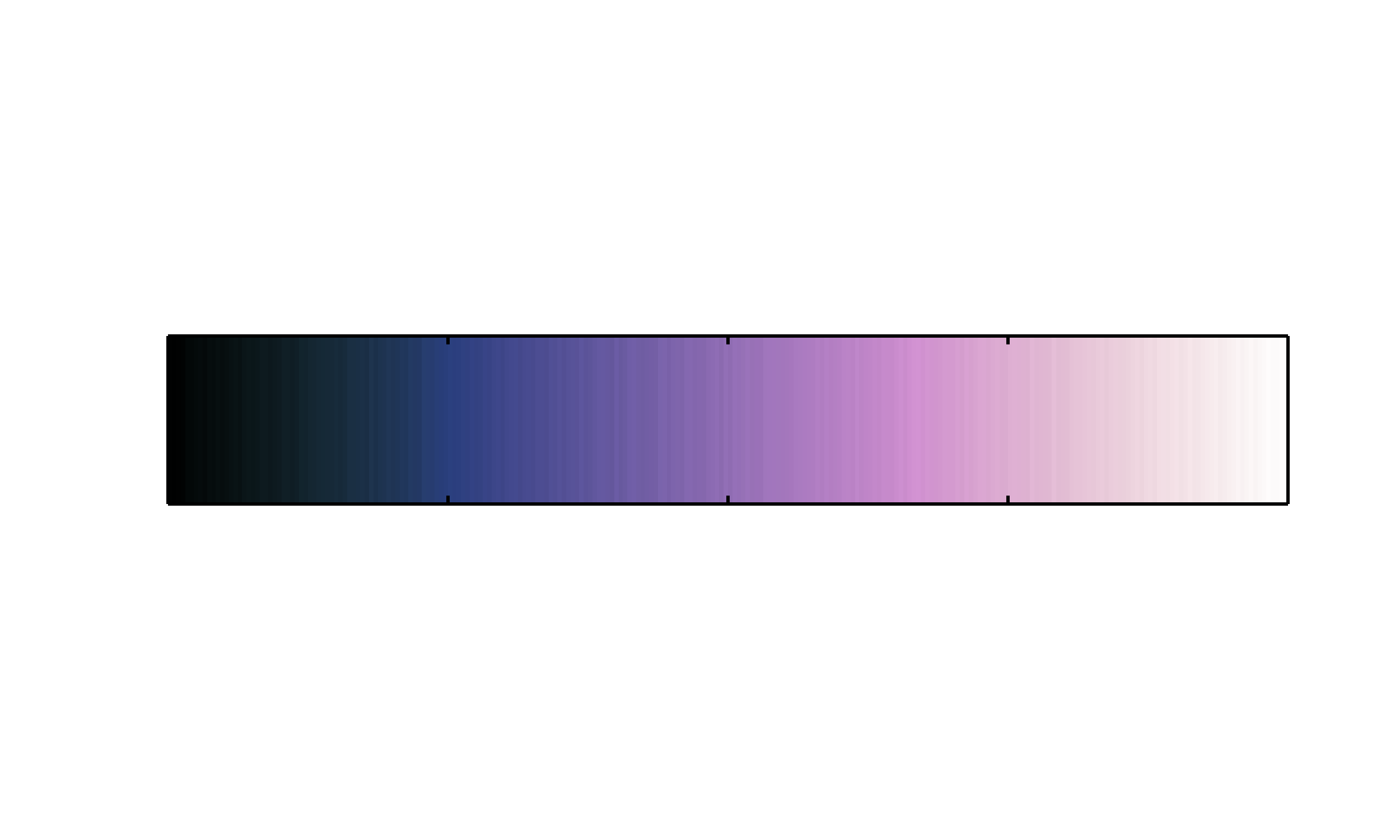}}}}
\caption{Maps of IGM gas temperature in CDM (left panels) and ETHOS
(middle panels). The image intensity shows squared gas density (with arbitrary normalisation), and the colour shows the temperature:
$\le10^{4}\,{\rm K}$ gas is shown in purple, $10^{5}\,{\rm K}$ in green and $\ge10^{6}\,{\rm K}$ in
red. The top panels show the entire box at $z=11$,
the middle two panels a zoom-in of the region highlighted with a white box in
the top middle panel at the same redshift, and the bottom panels show the same
zoomed region at $z=6$. Each image slice is $400\kpc$ thick; all lengths scales quoted
are comoving. In the right-hand panels we show the difference
map between the temperature of the CDM and ETHOS maps. Lighter regions are
hotter in ETHOS and darker regions are hotter in CDM (colour
bar in the bottom right panel).}
\label{Proj}
\end{figure*} 

We note that our simulations do not include radiative transfer and we
therefore cannot study reionization in a fully self-consistent framework. In
fact, our simulations are set up with a spatially uniform, time-dependent UV
background, which is not coupled to local star formation \citep[this is a
standard procedure when radiative transfer is not included, and the particular implementation we used is described in
Section 2.4 of][]{Vogelsberger13}. The simulations presented here follow the gradual build up at high redshifts of the UV background as prescribed by \citet{FG09}. Our approach is based instead on computing
the reionization history {\it a posteriori} using as input the star formation rate
density in our simulations. Our purpose here is to present a first-order
approximation of the expected relative differences between the reionization
history of CDM and the ETHOS benchmark model. We keep this limitation in mind when interpreting our results and note that all other cosmological hydrodynamical simulations that do not self-consistently model radiative transfer are subject to the same limitation. 

\section{Results}
\label{res}

Our goal is to study two high-redshift observables that could be used
to distinguish ETHOS models from CDM. The first is the abundance of galaxies at high
redshift, which is expected to differ towards lower masses due to the
primordial cut-off in the power spectrum. 
Second, changes in the high-redshift galaxy population will also
impact the UV photon budget, thereby changing the details of the reionization history in
both models. We note that although the first effect implies the second one, 
there are various compensating effects that can invalidate the simple
argument in favour of an overall reduced UV photon budget in ETHOS at high
redshift. We further note that the two effects are mostly caused
by the damping in the power spectrum and are not expected to be influenced
significantly by dark matter self-interactions, which are mainly responsible for
shaping the inner regions of collapsed haloes towards lower redshifts.   

We start our exploration of the high-redshift model differences with
Fig.~\ref{Proj}, where we give a visual impression of our simulations through a
series of maps of the intergalactic medium (IGM) that qualitatively show the
differences between the two dark matter models. We include maps for the entire
simulation volume and also a subregion of the volume that is host to a
star-forming overdensity.  The large-scale maps are presented at $z=11$, and
the zoomed region at both $z=11$ and $z=6$.  

The large scale structure of filaments, nodes, and voids is identical in the
two models, which is not surprising since ETHOS models 
preserve the large-scale clustering
characteristics of CDM. Differences become visible only at the smallest
resolvable scales, at which we find in general more structure in CDM than in
ETHOS due to the primordial damping of the power spectrum. These differences
become more apparent in the zoomed maps and their accompanying difference maps.
Here we also find that the heating of gas by galaxies is different in the two
models. The number density of heating sites is lower in the ETHOS model; the
small black regions in the difference maps show where a halo has collapsed in
CDM but not in ETHOS.  Furthermore, the expanding bubbles, driven by stellar
feedback, are in general smaller in ETHOS, since they formed later than in CDM
and had less time to expand. This phenomenon can be appreciated even better in
the difference maps (right panels of Fig.~\ref{Proj}): the regions that are
hotter in ETHOS (white) are often located within hot shells in CDM (black),
thus showing how the feedback bubble generation in ETHOS lags behind that of
CDM. 

\subsection{Abundance of galaxies at high redshift}\label{sec3_1}

The primordial power spectrum cut-off impacts the low-mass end of the ETHOS
halo-mass/luminosity function.  We investigate this effect quantitatively by
measuring the abundance of dark matter haloes and the $U$-band luminosity
function of the galaxies they host at $z\geq6$ in both models. We note that
ongoing star formation in galaxies is traced more directly with the intrinsic far UV (FUV)
luminosities, corresponding approximately to a rest frame wavelength of
$150\,{\rm nm}$, than by the $U$-band we use here, centred at around $365\,{\rm
nm}$. We study the FUV luminosity function in detail further below when we
show predictions for JWST (see section \ref{sec_JWST}).

We remark that, for simplicity, throughout this work we refer to {\it intrinsic} luminosities (absolute magnitudes) computed in a given band/wavelength without accounting for dust attenuation. The procedure to compute FUV luminosities from the simulation data is described further below in section \ref{sec_JWST}. 
For the other bands/wavelengths used in this work, the procedure is similar. 
Since we are avoiding the complication of dust modelling, we are also not properly taking into account the observational consequences that dust has in suppressing intrinsic luminosities, in particular the relevance it has in the observed $U$-band magnitudes of galaxies. Since this suppression is connected to the star formation history in a given galaxy, it will likely be different in ETHOS than in CDM. As we mention elsewhere, we are deferring the full and more detailed analysis of the properties of the ETHOS galaxies near the cut-off of the power spectrum for a future work.

\subsubsection{Impact on the halo mass function}

\begin{figure}
\includegraphics[width=0.49\textwidth]{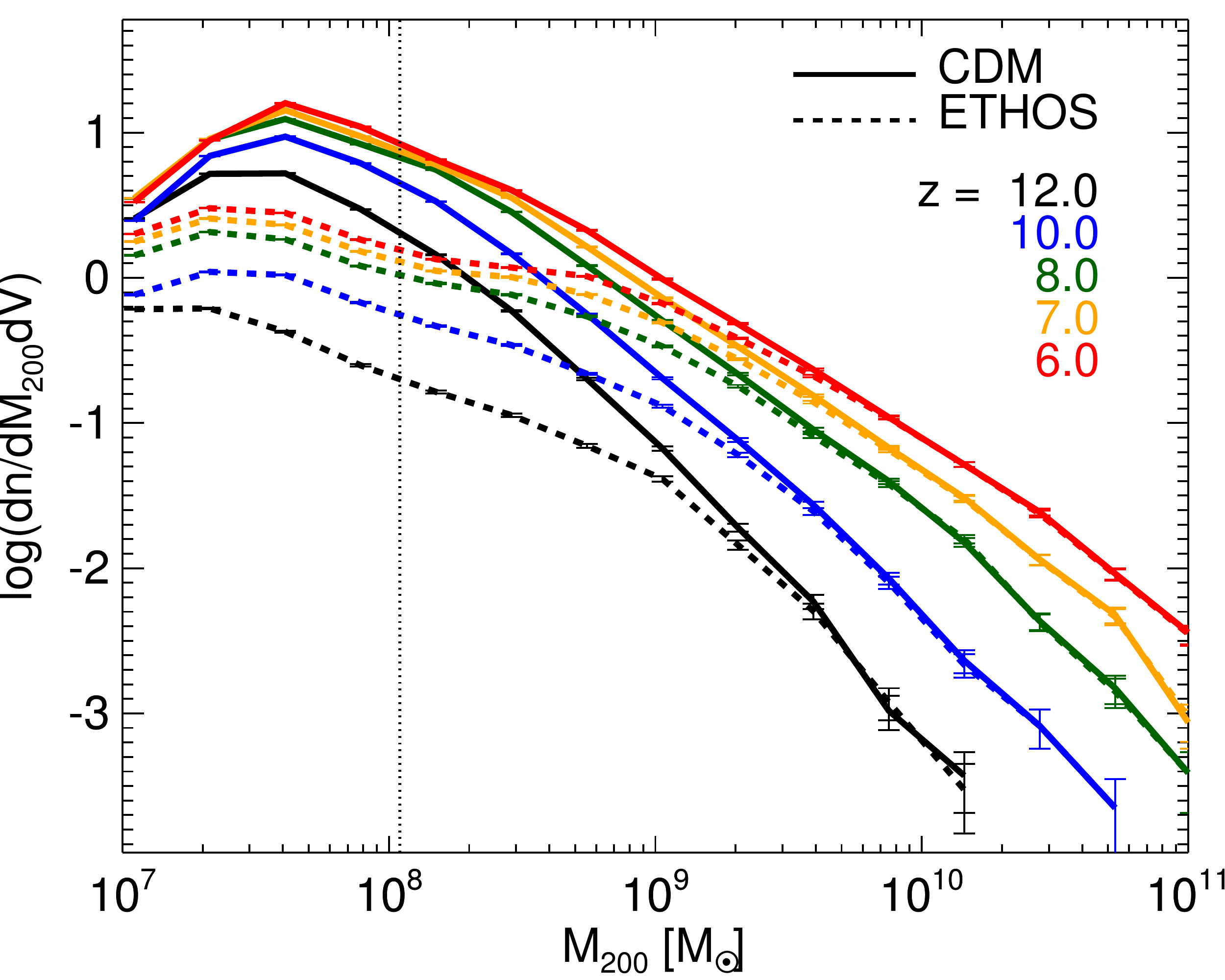}
\caption{Halo mass functions for CDM (solid lines) and ETHOS
(dashed lines) at five redshifts: z=[12,10,8,7,6] shown in black, blue,
green, yellow and red respectively. The error bars are Poissonian. The vertical dotted line is the effective halo mass resolution we use for both models, corresponding to the appearance of spurious haloes in ETHOS.}
\label{MhUV1}
\end{figure}

To start, we show the mass
function of haloes at redshifts in the range $z=[12,6]$ in Fig.~\ref{MhUV1}. We use the radius enclosing $200$ times the critical density to define the halo mass, $M_{200}$.
Both CDM and ETHOS show a strong cut-off at around $2-3\times10^{7}\msun$, which is the
resolution limit for the CDM case corresponding to $\sim20$ particles. There is
also a suppression in the ETHOS halo abundance compared to CDM by up to a
factor of several visible at that mass scale. However, these scales are clearly
affected by spurious fragmentation: there is a clear upturn in the mass
function below $\sim10^8\msun$, which agrees with the limiting mass due to
discreteness effects mentioned in Section \ref{sec:sims}. This mass scale can
be considered as the effective halo mass resolution for the ETHOS case. The
difference in the abundance of haloes at this resolved scale relative to CDM is
clearly apparent (a factor of $\sim5.5$ at $z=6$).  This would naively suggest that the
abundance of galaxies inhabiting these haloes, and thus the production rate of
ionizing UV photons, is suppressed in ETHOS relative to CDM by a similar
factor.  However, a halo can only act as a source of ionizing photons if the
gas it accretes can radiatively cool and collapse to form a luminous galaxy. If
the proportion of haloes that host galaxies -- the so-called luminous fraction
-- is different between CDM and ETHOS, then this effect will be relevant for the
production rate of ionizing photons in galaxies. 

\begin{figure}
\includegraphics[width=0.49\textwidth]{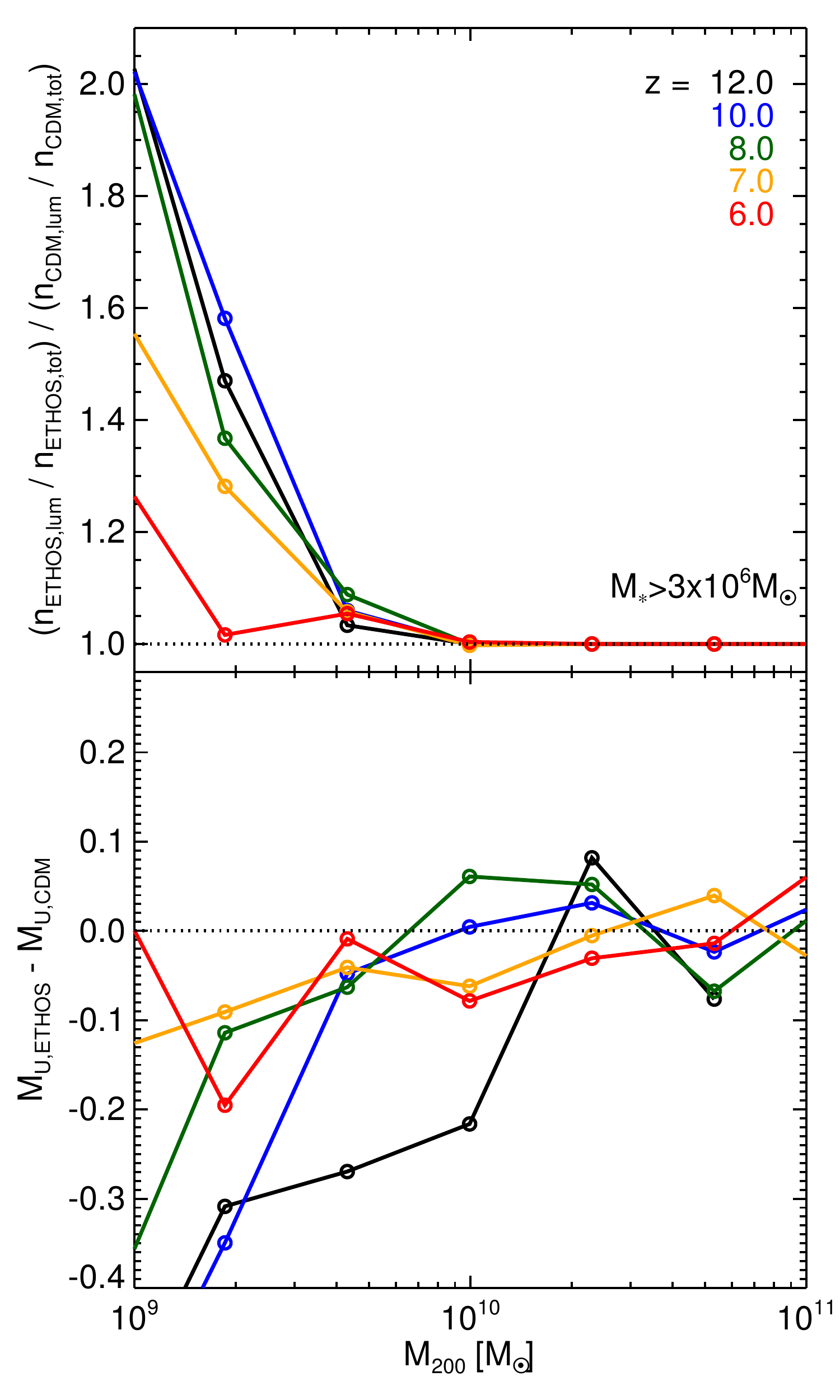}
\caption{Upper panel: Ratio of the fraction of haloes hosting galaxies with $M_\ast\geq3\times10^6\msun$ in ETHOS to that of CDM as a function of halo mass.
Lower panel: Difference in the median $U$-band magnitude between ETHOS and CDM
as a function of $M_\rmn{200}$.}
\label{MhUV23}
\end{figure}   

We therefore plot, in the top panel of Fig.~\ref{MhUV23}, the ratio of ETHOS and
CDM luminous fractions as a function of halo mass, where the luminous fraction
is defined as the fraction of haloes that contain a stellar mass larger than
$3\times10^6\msun$. We choose this stellar mass threshold to avoid spurious effects due
to limited resolution, and also restrict our plot to haloes of mass $M_{200}\ge10^{9}\msun$ as the galaxies hosted in $M_{200}<10^{9}\msun$ haloes rarely meet the stellar mass threshold, and are therefore subject to shot noise; note that we therefore resolve the haloes of all galaxies with this stellar mass. The plot demonstrates that below the halo mass where the
halo mass function in ETHOS starts to deviate from CDM, the luminous fraction
is actually higher in ETHOS than in CDM. The difference grows towards lower
masses and becomes substantial below the halo resolution limit,
although the statistics of haloes hosting galaxies with stellar mass beyond the
chosen threshold are poor in this limit. We have found that lowering the stellar mass threshold eases the difference between the two models somewhat, which emphasizes the fact that although $<10^{10}\msun$ CDM haloes do host galaxies, they are less massive than their ETHOS counterparts. This plot indicates that despite their lower number, ETHOS haloes with scales near the primordial cut-off in the power spectrum have a higher star formation efficiency than their CDM counterparts. We speculate that this higher efficiency is the
result of an enhancement in starbursts, as the first haloes in ETHOS form
through a monolithic collapse and not hierarchically. As these gas-rich haloes
merge, they produce brighter starbursts than in the CDM case. A similar
phenomenon has been described in the WDM context \citep{Bose16c,Bose17a}, as we
discuss below. 

There might be an environmental effect as well linked to this phenomenon. In ETHOS, galaxies within haloes of masses between
$(10^8-2\times10^9)\msun$ are the first to form, with no prior star formation in
less massive haloes. Thus, although there is a dearth of haloes in this mass range,
caused by the primordial power spectrum cut-off, there is also a 
compensating effect since the absent haloes in ETHOS are clearly 
not a source of (stellar) feedback into the local environment
around them, as they are in the CDM case. Visually, this can be appreciated in 
the right panels of Fig.~\ref{Proj} by the absence in ETHOS of the galactic wind 
bubbles driven by stellar feedback  within the smallest haloes seen in CDM. Thus, we speculate that 
star formation
within these smallest haloes is an additional source of heating (through stellar
feedback) of the local environment, which might suppress star formation within 
nearby larger haloes. These same haloes would be unaffected in ETHOS, where 
this source of heating is absent. Fig.~\ref{Proj}
provides a degree of qualitative evidence for this speculative mechanism, since the extent
and density of feedback-driven bubbles is seemingly larger in the CDM case (bottom-right
panel of Fig.~\ref{Proj}). That these bubbles affect more strongly other nearby haloes in CDM than
in ETHOS seems plausible. 

The interplay between the underabundance of low-mass haloes and the delay of
the onset of galaxy formation has been studied in the context of WDM,
which has a primordial power spectrum cut-off similar to the one in ETHOS.
\citet{Bose16c,Bose17a} showed that, when implementing their semi-analytic
model of galaxy formation in a WDM cosmology, galaxy formation is indeed
delayed, but the first galaxies that form in WDM are more massive and more gas
rich than their CDM counterparts at a fixed halo mass. This results in brighter
starbursts, i.e., high star formation rates, as these galaxies form.
Therefore, at high redshift, the formation of bright starbursts is more
efficient in WDM than in CDM, which also leads to a larger number of ionizing
UV photons in WDM compared to CDM. The interplay between these two effects
depends on the details of the galaxy formation model and the scale where the
primordial cut-off of the power spectrum happens. Under certain conditions, the
enhancement of earlier bright starbursts might be efficient enough to produce a
UV luminosity function with a higher amplitude in WDM than in CDM, as it was
found in \citet{Bose17a} for a wide range of UV luminosities for $z>5$. The difference in that case was stronger at higher redshifts across all masses,
while at a fixed redshift, the difference was larger for more luminous
galaxies. For the faintest galaxies the trend was actually reversed, with the
amplitude being higher in CDM (e.g., at $z=7$, this reversal happens at $M_{\rm
AB}(UV)\sim-12$; see Fig. 12 of \citealt{Bose17a}). On the other hand, 
\citet{Bose16c} explored WDM models similar to those in
\citet{Bose17a} but with a different supernova feedback implementation
(seemingly more consistent with lower mass galaxies at $z\sim0$) and found that
the UV luminosity function in WDM is always below the CDM case (see Fig. 6 of
\citealt{Bose16c}). 

\subsubsection{Impact on galaxy luminosity functions}

We investigate the interplay between these competing effects in our
simulations. We start by computing the median $U$-band luminosities at each
halo mass and plot the difference in the median between ETHOS and CDM as a function of
$M_\rmn{200}$ in the lower panel of Fig.~\ref{MhUV23}. For all redshift-halo-mass combinations at which
we have good statistics, there is a clear preference for ETHOS galaxies to be
brighter than CDM ones for host haloes with masses $<10^{10}\msun$. In this respect, our results are qualitatively similar to those in
\citet{Bose16c,Bose17a}.

There are interesting features in the behaviour with redshift of the competing scales
that set the galaxy formation threshold. Baryonic physics (mainly heating from reionization) suppresses galaxy formation
for halo masses below $10^{10}\msun$\footnote{Strictly, for our purposes, this mass scale is defined as the halo mass where the luminous fraction of haloes (i.e. those having galaxies with $M_\ast\geq3\times10^6\msun$), starts to be less than 1.}, with this mass threshold increasing at
lower redshifts as reionization feedback inhibits star formation in progressively
larger objects at increasingly lower redshifts (see e.g.
\citealt{Okamoto08,Sawala16b}). This hierarchy is inverted when comparing the
suppression (driven by new dark matter physics) of ETHOS haloes relative to CDM haloes where the mass threshold for suppressing halo formation
becomes smaller with decreasing redshift. This can be understood as the
transfer of power from large to small scales, which causes the evolution of the
power spectrum to `catch up' with CDM (clearly reflected in the halo mass
function in Fig.~\ref{MhUV1}).

We have found that the two mass scales where galaxy formation is
suppressed by either dark or baryonic physics are quite similar:
$M_{200}\approx10^{10}\msun$. This coincidence is driven by the arbitrary choice of stellar
mass threshold used to define the luminous fraction
($M_\ast=3\times10^6\msun$), but it nevertheless illustrates a relevant point:
if a downward change in the slope of the luminosity function were to be
detected towards low luminosities, distinguishing it from a primordial cut-off
in the power spectrum would be challenging. By contrast, the inverted behaviour
of the mass threshold for galaxy formation with redshift resulting from dark
and baryonic causes is a promising signature to look for in upcoming
observations. Ultimately, this is a first order analysis, a more detailed
examination of this process from the theoretical perspective will require
significantly higher numerical resolution to map the cut-off in detail ($\sim6\times10^{4}\msun$ in dark matter particle mass to resolve haloes near the cut-off with $\sim$100 particles), an in-depth
exploration of the synergy between a primordial cut-off in the power spectrum,
and different implementations of the physics responsible for reionization, ideally
through direct radiative transfer calculations.  

\begin{figure}
\includegraphics[width=0.49\textwidth]{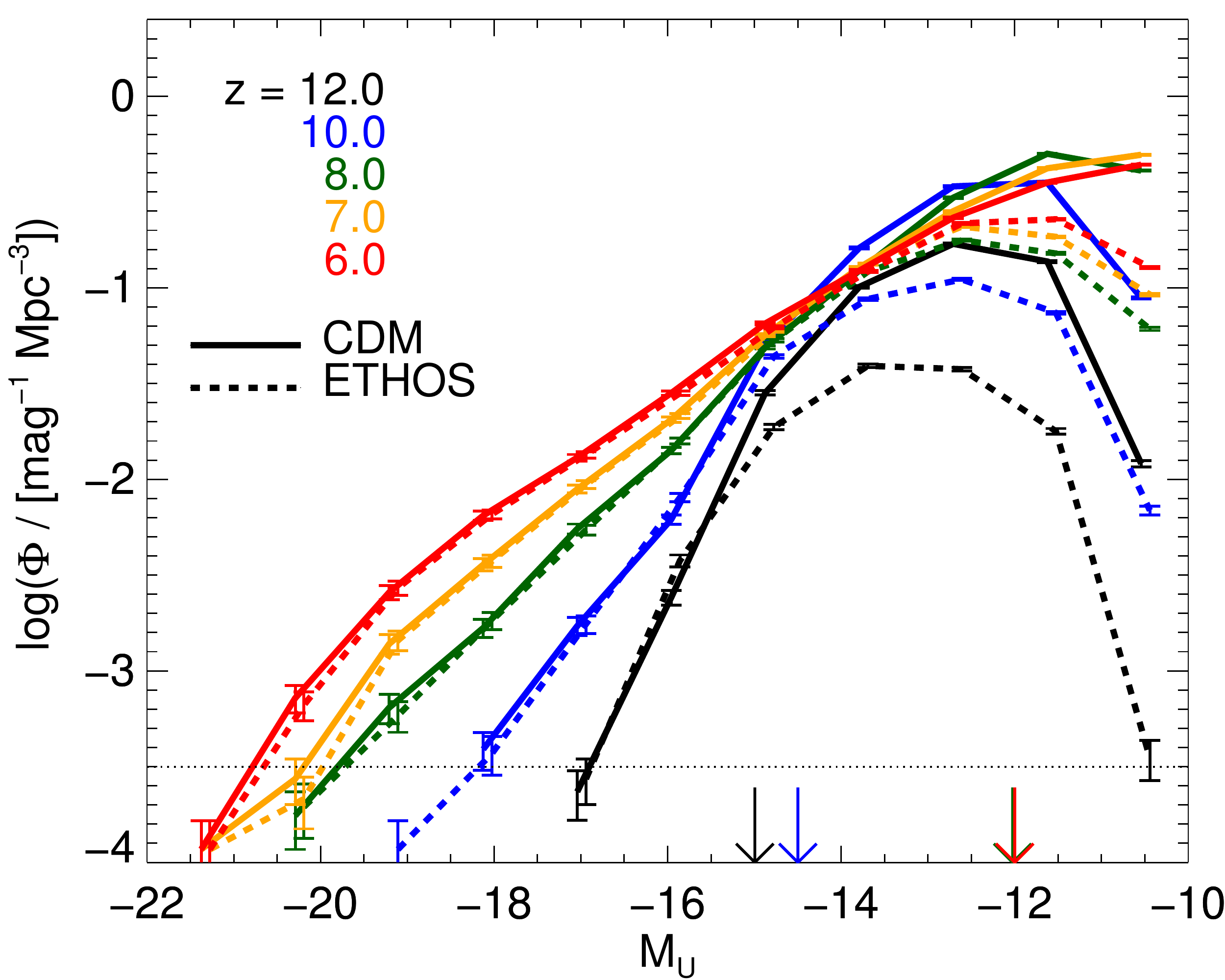}
\caption{The $U$-band luminosity functions for CDM (solid curves) and ETHOS
(dashed curves). Colours are for different redshifts $z=12$, $10$, $8$, $7$ and $6$ for
black, blue, green, orange, and red respectively. The horizontal dotted line marks the galaxy abundance below which low number statistics in the simulation affect the results in a relevant way ($<16$ galaxies per bin). The coloured arrows in the horizontal axis mark the approximate magnitude at each redshift below which a fraction of galaxies (approximately 16$\%$, i.e. the fraction below the lower $1\sigma$ region of the distribution of galaxies in the $M_{200}-M_U$ plane) are hosted by $10^{8}\msun$ haloes, and therefore the luminosity functions are at least partially suppressed by mass resolution.}
\label{UVFunc}
\end{figure}  

The rate of ionizing photon production is ultimately a convolution of the halo
mass function with the UV luminosity per halo mass. The suppression in the ETHOS halo mass function will lead to a lower
number of galaxies in total compared to CDM, but the higher luminosity per halo
may perceptibly lead to an enhancement in the relative number of bright
galaxies, for some threshold in luminosity. To check to what degree either of
these is the case, in Fig.~\ref{UVFunc} we plot the $U$-band luminosity
functions for our two models at five redshifts in the range $[12,6]$.      

At all redshifts, the most marginally resolved galaxies are suppressed in ETHOS
relative to CDM with a gap that closes for lower redshifts. There is no
redshift at which the abundance of bright ETHOS galaxies exceeds (in a
statistically significant way) that of CDM, therefore the enhanced $U$-band
luminosities of ETHOS galaxies succeeds only in diminishing the intrinsic
difference between the ETHOS and CDM halo mass functions. In this regard, and
in comparison with WDM, our results are close to those of \citet{Bose16c}, and
thus, we do not find the overabundance of bright galaxies relative to the CDM case
as reported in \cite{Bose17a}.

The gap between ETHOS and CDM closes almost completely by $z=6$\footnote{This is strictly valid at our resolution limit of $M_U\sim-12$ for $z>8$. We notice that at higher redshifts, the underabundance of low-mass haloes, and thus low-mass galaxies is strong in ETHOS, but at $M_U\sim-12$, resolution issues are relevant at higher redshifts (see arrows in the horizontal axis in Fig.~\ref{UVFunc}) and is thus not possible to quantify the effect adequately.}, which implies that the galaxy population builds up more
rapidly in the ETHOS model than in CDM ( similar to what was found in  \citealt{Bose16c} for WDM, although \citealp{VillanuevaDomingo18} found a more persistent difference.). This rapid
buildup is a generic feature of models with a cut-off in the power spectrum, seemingly irrespective of the details of the
galaxy formation model. This points to a promising observational feature to look for in future observations at the low end of
the high-redshift luminosity function.

We note that the strong drop off in the abundance of low-luminosity galaxies at
the highest redshifts ($z>8$) in Fig.~\ref{UVFunc} is driven at least in part by the resolution
limit of our simulations. This is an effect caused by galaxies having higher
ongoing star formation rates, and thus being brighter (for a given halo mass),
at high redshifts. Thus, at a fixed ($U$-band) magnitude, the haloes hosting
these galaxies have progressively lower masses at larger redshifts.  Once the
typical halo mass reaches our resolution limit for the halo mass function
($\sim10^8\msun$), the abundance of haloes, and hence of galaxies of the
associated magnitude, starts being artificially suppressed. For $z\leq8$, this
is not an issue down to $M_U=-12$ since the median halo mass at that magnitude
is $\gtrsim6\times10^8\msun$. The flattening of the U-band luminosity function
towards lower magnitudes at lower redshift is thus a resolved feature in our
simulations and is due to the lower star formation efficiency at lower redshift
driven by feedback (stellar and ionizing background). For $z\leq8$ we are thus
confident that the $U$-band luminosity function is sufficiently resolved down
to $M_U=-12$\footnote{We note that we have verified, with a lower resolution
set of simulations (by a factor of 8 in mass resolution), that the $U$-band
luminosity function is converged, in the low resolution case, down to the
magnitude corresponding to the typical halo mass where the halo mass function
is converged.}. For $z>8$, the $U$-band is progressively more affected by
resolution and by $z=14$, it is properly resolved down to $M_U\sim-15$ only (see arrows in the horizontal axis in Fig.~\ref{UVFunc}). It is possible that this cut-off may be partly physical if there is a minimum luminosity associated with the initial starburst; we will examine this possibility in a future paper.

\subsubsection{Predictions for JWST}\label{sec_JWST}

\begin{figure}
\includegraphics[width=0.44\textwidth]{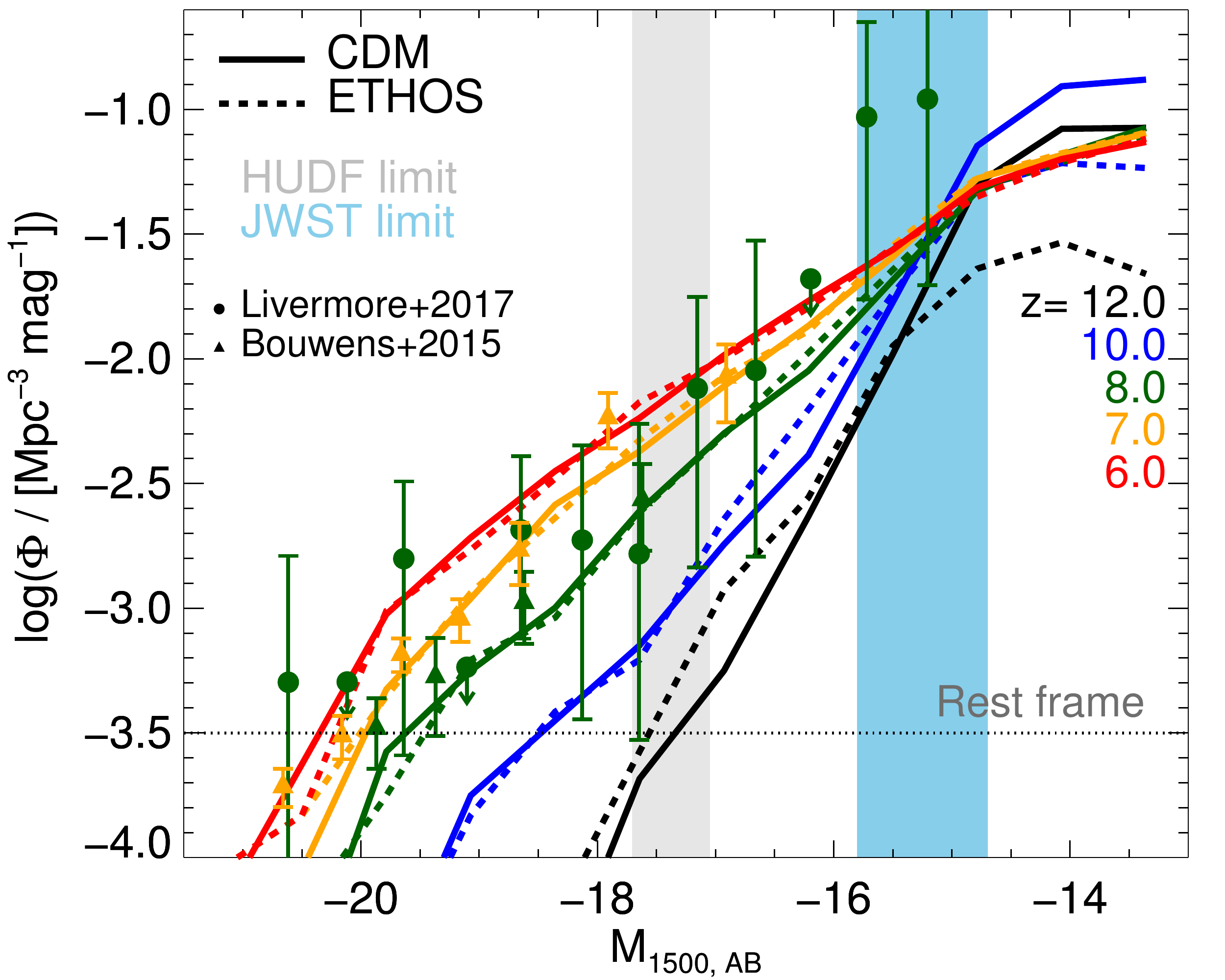}
\includegraphics[width=0.44\textwidth]{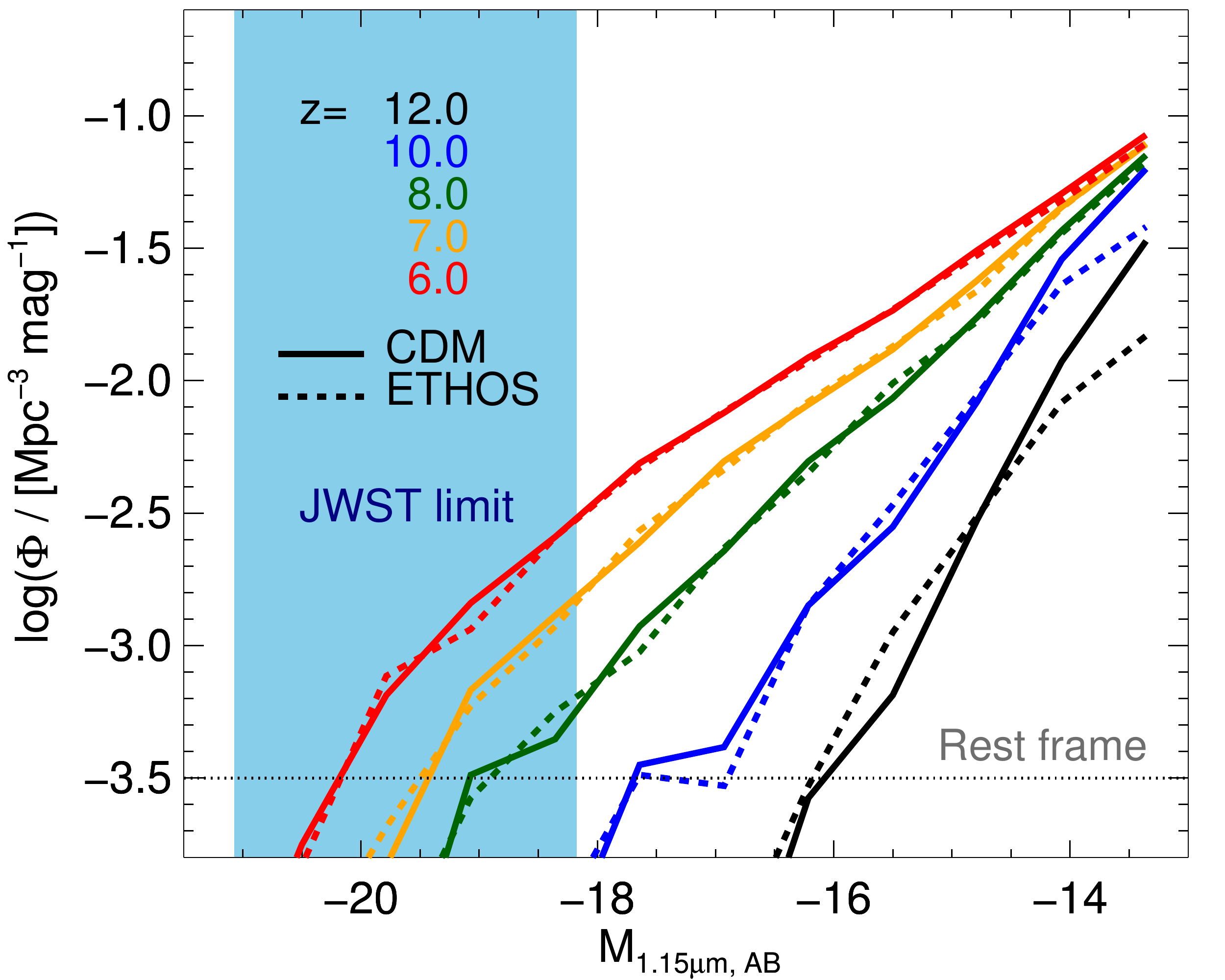}
\includegraphics[width=0.44\textwidth]{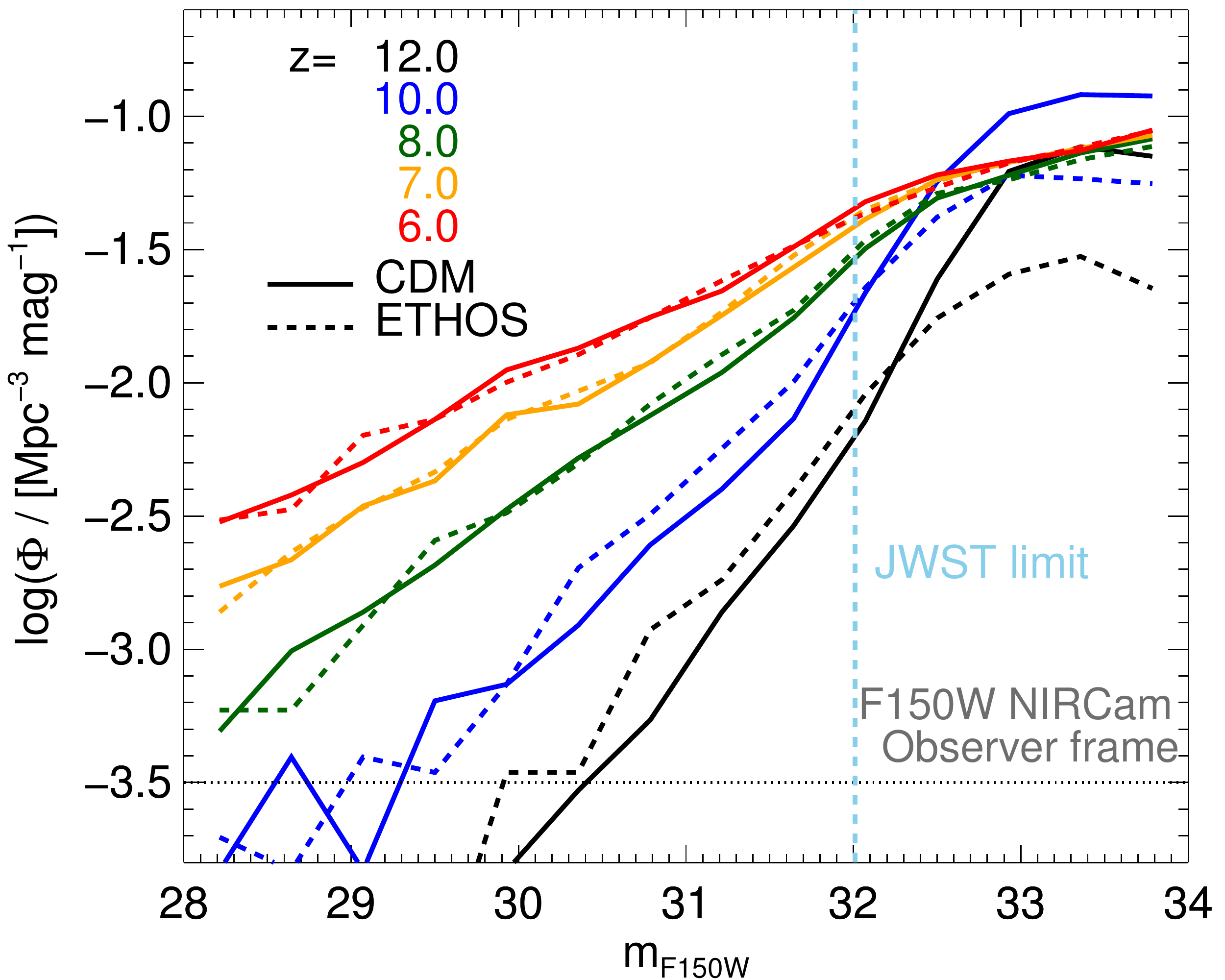}
\caption{FUV (150~nm) and NIR (1.15~$\mu$m) rest-frame luminosity
functions, on the top and middle panels, respectively, plus the 
luminosity function in the observer-frame (using apparent magnitudes)
in the JWST F150W band in the bottom panel. The different colours
are for different redshifts according to the legend, and the solid and dashed
lines are for the CDM and ETHOS cases, respectively. The horizontal dotted line marks the galaxy abundance below which low number statistics in the simulation affect the results in a relevant way ($<16$ galaxies per bin). For the upper panel a collection of observations is also shown
\citep{Bouwens15b,Livermore17}. The grey (top panel) and blue (top and middle panels) bands are estimated observational limits from HUDF and for an optimistic deep
survey with JWST. Interestingly, the differences between
ETHOS and CDM start just to be observable at the limit
of JWST. In the bottom panel we show the expected JWST magnitude limit in the observer-frame for the F150W NIRCam filter.}
\label{JWST_LF}
\end{figure}  

We have demonstrated so far that the galaxy populations in CDM and ETHOS behave
differently at high redshifts near the primordial power spectrum cut-off. A key question is then whether such a difference can be detected to distinguish these models observationally. To this end, we
compute the luminosity function in our simulations at wavelengths that will be observed by  
JWST. We do this using the Flexible Stellar Population Synthesis (FSPS) code\footnote{https://github.com/cconroy20/fsps} \citep{Conroy09,Conroy10}. For each star particle in a given simulated galaxy, we construct a simple stellar population (SSP) using as input the metallicity and age of the star particle, and using the initial mass function (IMF) used in our simulation setting \citep[Chabrier IMF;][]{Chabrier03}; the code then outputs the spectra of the SSP for the particle. A mass-weighted sum is then performed across all particles in the galaxy to compute its spectral energy distribution and total luminosity in the desired band. We compute the FUV and Near Infrared (NIR) luminosity functions, at 150 nm and 1.15 $\mu$m rest frame wavelengths, top and middle panels of Fig.~\ref{JWST_LF}, respectively. We choose these two wavelengths since they are representative of the FUV, which is a good tracer of recent star formation (young stars), and the NIR, which is a better tracer of the older stellar population (more sensitive to the prior star formation history). In the bottom panel of Fig.~\ref{UVFunc} we also present the evolution of the luminosity function (in the observer frame) as it would be observed by the Near InfraRed Camera (NIRCam) on JWST (filter F150W), taking into account the transmittance of the NIRCam Filter in JWST\footnote{https://jwst-docs.stsci.edu/display/JTI/NIRCam+Filters}.

The luminosity functions in Fig.~\ref{JWST_LF} are shown in monochromatic AB magnitudes, rest-frame in the upper and middle panel, observer-frame in the bottom panel.
The FUV (150 nm) luminosity function is shown in the upper panel of Fig.~\ref{JWST_LF}. The grey vertical
band is roughly the current limit from HST observations (HUDF and CANDELS, see e.g.
\citealt{Bouwens15b}\footnote{ The \citet{Bouwens15b} results were measured at 160~nm rather than 150~nm; we expect that this difference does not affect our conclusions.}),
while the blue band is the estimated limit for JWST, which is based on the sensitivities for the NIRCam for point source detection with a signal to noise ratio ($S/N$) of 10 and $10^4$~s exposure\footnote{The F115W, F150W and F200W are the NIRCam filters sensitive to the rest-frame FUV (150nm) luminosity function in the redshifts shown in the top panel of Fig.~\ref{JWST_LF}. Their sensitivities were taken from https://jwst.stsci.edu/instrumentation/nircam}. We scaled these sensitivities for the fairly optimistic scenario of a deep field survey with $10^6$~s exposure (assuming a $t^{-2}$ scaling), a factor of a few better than the HUDF, and lowering the threshold for point source detection to $S/N=5$. The limit is shown as a band, since the flux sensitivities in Jy are transformed into redshift-dependent sensitivities in the rest-frame magnitudes. We observe that it is approximately at the limit of what JWST can observe in the FUV where the difference between CDM and ETHOS starts to be apparent. Unless
the actual final survey strategy and depth for JWST is improved, it will be
difficult to distinguish the models in this way, albeit the high-redshift
range $z=10-12$ might be promising. 

The rest-frame NIR (1.15$\mu$m) luminosity function for our simulations is shown in the middle panel of
Fig.~\ref{JWST_LF}. Since this wavelength is more sensitive to the older stellar
population, and hence to the star formation history, it becomes less sensitive,
particularly at higher redshifts, to the enhanced starburst phenomena in ETHOS
discussed earlier, which mostly affect the recent star formation in the galaxy.
The rapid build-up of the galaxy population at the fain-end observed in the FUV is thus not as apparent in the NIR. The difference between the ETHOS and CDM models is however, not apparent until $z\geq8$ for $M_{{\rm AB}}(1.15\mu{\rm m})=-14.5$.

The sensitivity of JWST to NIR wavelengths relies
on a different instrument, the Mid InfraRed Instrument (MIRI), which is considerable less sensitive than NIRCam. With a similar optimistic survey scenario as the one described above, we show the sensitivity limit of JWST for the NIR(1.15$\mu$m)\footnote{The F770W, F1000W, F1130W and F1280W are the MIRI filters sensitive to the rest-frame NIR (1.15$\mu$~m) luminosity function in the redshifts shown in the middle panel of Fig.~\ref{JWST_LF}. Their sensitivities are taken from: https://jwst.stsci.edu/instrumentation/miri} in the middle panel of Fig.~\ref{JWST_LF}. The prospects of JWST reaching the desired magnitudes in NIR are thus extremely low.  

A promising strategy is to use gravitational lensing to reach fainter
magnitudes. Using data from the Hubble Frontier Fields program, it has been possible to detect
very faint galaxies strongly lensed by galaxy clusters. This development makes
it possible to probe the UV luminosity function to very faint magnitudes
\citep{Livermore17,Ishigaki18}, close to $M_{UV}\sim-15$ between $z=7-9$.  
At these magnitudes and redshifts, the $68$~per~cent
confidence interval has an amplitude of $\sim1\,{\rm dex}$, which is a factor
of $\gtrsim10$ too large compared to the differences between CDM and the
benchmark ETHOS models\footnote{Notice that increasing the number of cluster
lenses would increase the effective volume of the lensing survey and thus would
reduce the statistical errors in the reconstructed luminosity function.}. As a reference, the observations from \citet{Livermore17} at $z=8$ have been added to the upper panel of Fig.~\ref{JWST_LF}.
With upcoming surveys with the JWST, the prospects of exploiting lensing
magnification in a similar way to constrain a primordial cut-off in the power
spectrum are promising when combined with a good understanding of the physics
of galaxy formation. Although challenging, we think that this might lead to
powerful high-redshift constraints for alternative dark matter models in the
near future.
 
\subsection{Impact on reionization}

Above we have studied the differences in the abundance of galaxies in ETHOS
and CDM, and pointed out that these differences are not detectable with existing instruments, but could potentially be revealed by upcoming telescopes like JWST. However, we can use
the predicted galaxy populations in both models to estimate the optical depth
for reionization in CDM and ETHOS. This can potentially constrain or rule out
certain non-CDM models, and the question we want to tackle here is whether our
benchmark ETHOS model is consistent with current measurements of the
reionization history encoded in the optical depth observations. We
demonstrate that this is indeed the case. 

As mentioned earlier, our simulations do not have radiative transfer and the UV
background used in them  is not coupled to the actual star formation.  Because
of this, our approach is based on estimating the fraction of gas that would be
ionised in the IGM due to star formation in the simulated galactic population.
In particular, we use the predicted star formation rate (SFR) density,
$\rho_\rmn{sfr}$, to determine the overall production rate of ionizing photons.
We thus need to ascertain this quantity as a function of redshift, taking into
account the fact that, in CDM at least, it is the faintest, and thus most
numerous, galaxies that generate the bulk of the re-ionizing photons.

\begin{figure}
\includegraphics[width=0.49\textwidth]{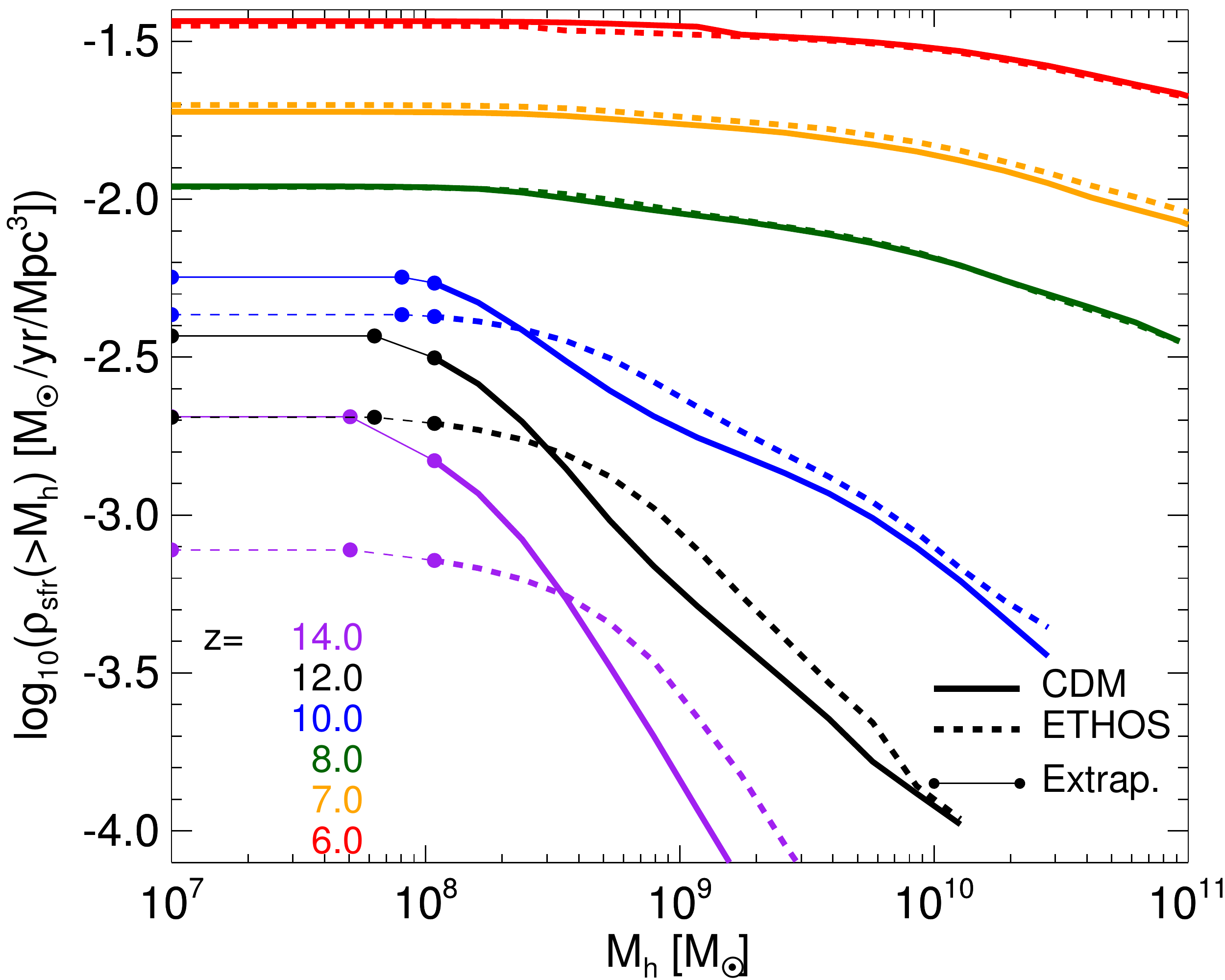}\\
\includegraphics[width=0.49\textwidth]{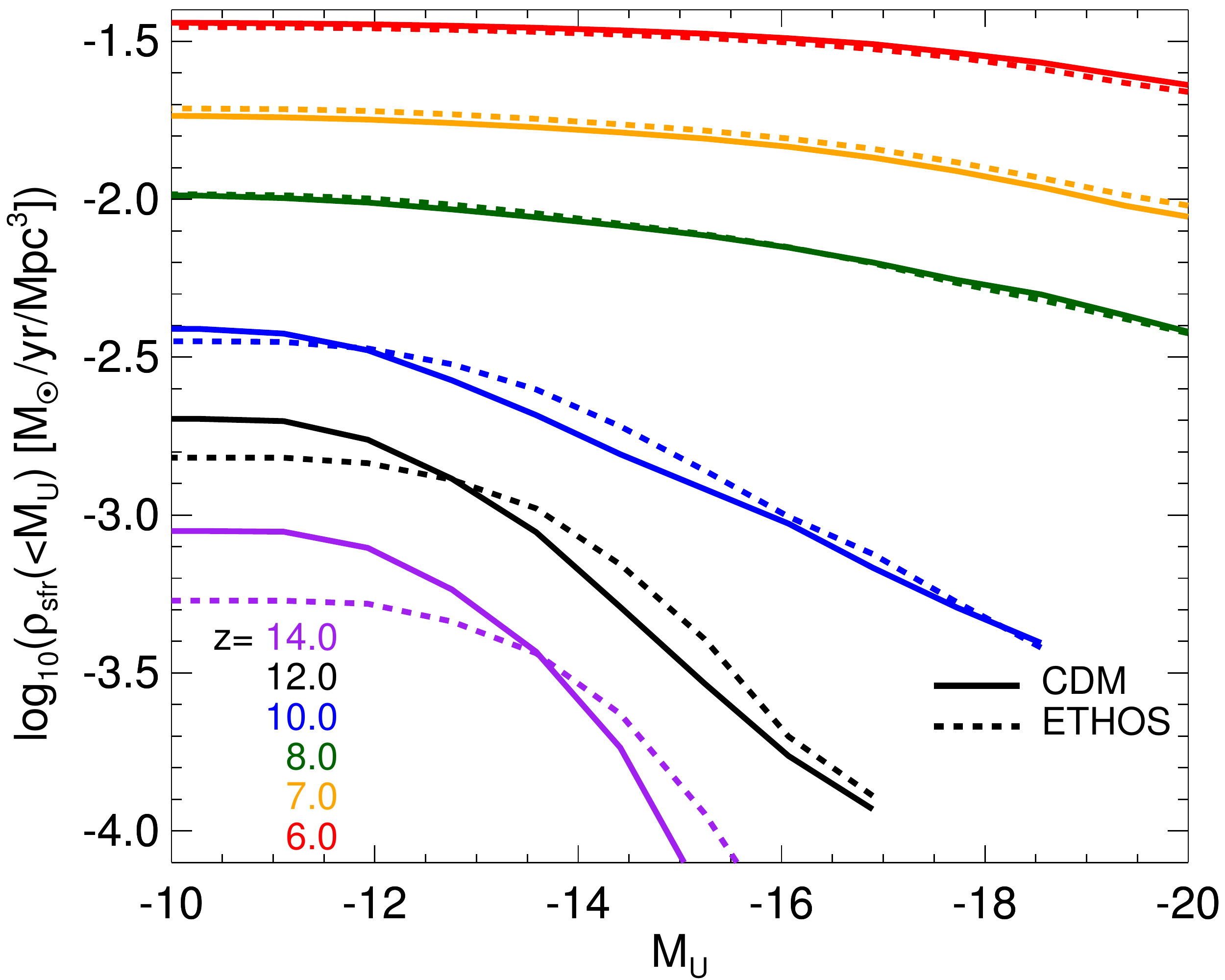}
\caption{Cumulative SFR density as a function of the halo mass, $M_\rmn{h}$ (top panel) and $U$-band luminosity $M_{U}$ (bottom panel). Solid (dashed) lines are for CDM (ETHOS). Simulation data
is show as thick lines, extrapolations below $M_{200}=10^{8}\msun$ are shown as thin lines, with the two solid circles (top panel only) bracketing the range where the extrapolation was used. Different colours are for different redshifts according to the legend. Note the addition of the redshift $z=14$ (purple) over  previous plots.}

\label{UVrhoExp}
\end{figure}

These faint galaxies are constrained to inhabit haloes that are sufficiently massive for gas to cool to high enough densities to form stars. We take this limit to be given by the virial temperature of the halo at which primordial gas can cool via atomic transitions: $T_{\rm vir}=10^4~$K. The corresponding mass limit is in the range $5\times 10^{7}-1.6\times10^{8}\msun$, in the range $6\leq z\leq 14$, with higher mass thresholds for lower redshift. This limit is tantalisingly close to the mass resolution of our simulations. Even though these low mass haloes have poor resolution in star particles (or even are devoid of star particles), the imposed star formation equation of state enables us to calculate the expected SFR given the gas content in a halo, independent of whether this halo contains any star particles. Therefore, whereas other studies have relied on recipes to calculate the UV luminosity of galaxies, we are able to relax many of the assumptions involved in this process by using the SFR in a halo given directly from the simulation data. We thus plot the cumulative SFR as a function of halo mass, $M_\rmn{h}$, in Fig.~\ref{UVrhoExp}, where our definition of halo mass is here the gravitationally bound mass ascribed by the {\sc subfind} algorithm in order to include subhaloes as well as host haloes. We also include an extrapolation between the spurious-halo resolution threshold $10^{8}\msun$ and our computed cooling limit for the halo mass for those redshifts at which the cooling limit is below the resolution limit.

We find that the extrapolation introduces very little extra additional star formation as predicted. The effect of the
extrapolation is negligible at low redshift due to the flattening in the
$M_\rmn{h}-\rho_\rmn{sfr}$ relation. The faint slope is shallower in ETHOS, thus
the additional star formation introduced by the extrapolation is less than in
CDM. CDM clearly produces more stars than ETHOS, despite the enhanced starburst nature of the brighter galaxies in ETHOS. 

The behaviour of the cumulative SFR with halo mass is very useful for performing these extrapolations in simulation-based studies, but it cannot be compared directly with observations. We therefore also include in the lower panel of Fig.~\ref{UVrhoExp} the cumulative SFR plotted as a function of the $U$-band magnitude. The qualitative behaviour of this relation mirrors that of the halo mass counterpart, with brighter ETHOS galaxies exhibiting higher SFRs only for the contribution of faint CDM galaxies to dominate the total budget. Therefore, the slope of the observed relation is a discriminant between the two models, notwithstanding the difficulties of making these observations as has been shown in the previous section. Note also that, at $z>10$, the total SFR as measured using $M_\rmn{h}$ is significantly higher than from the $U$-band plot. This is a resolution effect, since in the former case we can measure the SFR from the gas in $<10^9\msun$ haloes that host no star particles, and therefore do not have a measured $M_{U}$; this contribution would therefore be missed if we were to use $M_{U}$ for our measurement.

We have thus far assumed that all star formation occurs inside collapsed haloes, as these are the only regions in which gas cooling is efficient. However, it is possible that some gas cells in the simulation are not assigned accurately to haloes due to limited resolution, particularly when the host halo is still undergoing its initial collapse. Second, star formation could occur sporadically in uncollapsed regions, which could
potentially be relevant for the global SFR before resolved haloes have had an opportunity to collapse. Finally, in the ETHOS model there is also a physical case to be made that some star formation will occur outside haloes. It has been shown using very high resolution simulations (dark matter particle mass $\sim10^{2}\msun$) that the WDM cosmology generates smooth filaments that can attain gas densities high enough to form stars \citep{Gao07}. Since ETHOS models behave similarly to WDM, they may exhibit the same effect.

To check for the influence of star formation within unbound regions, we compute the total SFR density in all gas particles in our simulated volume, $\rho_\rmn{sfr,T}$, and the total SFR density in all gas particles that are bound to haloes, $\rho_\rmn{sfr,B}$; the difference between the two is the SFR density in unbound regions $\rho_\rmn{sfr,UB}$. 
The quantity of interest is the ratio of $\rho_\rmn{sfr,UB}$ to $\rho_\rmn{sfr,T}$, which is the contribution to SFR that occurs outside haloes. 

For CDM, this ratio is 25~per~cent at $z=20$, which drops to 4~per~cent by $z=14$, holds steady until $z=10$, and then it drops further to 1~per~cent at $z=6$. We speculate that the remarkably high fraction at $z>14$ is due to numerical resolution, which reduces to a percent level of the total when haloes start to collapse. The ETHOS simulation shows slightly higher unbound fractions for $z<10$, which is likely due in part to the delay in structure formation caused by the power spectrum cut-off. The discrepancy between the two models could also be explained by some combination of a lower total halo-based SFR in ETHOS or by extra SFR in the proposed filament mode, the latter of which could be manifest as an excess of unbound star formation in ETHOS compared to CDM. We have therefore checked the total unbound SFR in both models, and find that CDM obtains the higher unbound SFR at all redshifts, contrary to what we expect if filament star formation were relevant. We conclude that the excess star formation is due to background noise and resolution effects, and that its magnitude is sufficiently small (particularly for $z<14$) that it does not affect our results.

We can use our measurements and extrapolations of the SFR density to estimate the optical depth using the analytic procedure introduced
in \cite{Kuhlen12} (see also \citealt{Schultz14,Robertson15}). The procedure converts an
input star formation rate into an ionizing photon rate, which is then used
to calculate the optical depth:
\begin{equation}
\tau(z)=c\,\langle n_H\rangle\,\sigma_T\int_0^z f_e \, Q_{{\rm H_{II}}}(z')\,H^{-1}(z')\,(1+z')^2 dz',
\end{equation} 
where $c$ is the speed of light, $H(z)$ is the Hubble parameter, $\sigma_T$ is
the Thomson cross section, ${\langle n_H\rangle=X\,\Omega_b\,\rho_{\rm crit}}$ is the
comoving background density of hydrogen with $X=0.75$ being the hydrogen mass
fraction, and $\rho_{\rm crit}$ the critical density. The number of free
electrons per hydrogen nucleus is $f_e=1+\eta \, Y/4X$, where $Y=0.25$ is the
helium mass fraction and we consider helium to be singly ionised ($\eta=1$) at
$z>4$ and doubly ionised ($\eta=2$) at lower redshifts. The volume filling
fraction of ionised hydrogen $Q_{{\rm H_{II}}}$ is given by the differential
equation:
\begin{equation}
\frac{\rmn{d}Q_{{\rm H_{II}}}}{\rmn{d}t}=\left(\frac{1}{\langle n_H\rangle}\right)\frac{\rmn{d}n_{\rm ion}}{\rmn{d}t}-\frac{Q_{{\rm H_{II}}}}{t_{\rm rec}},
\end{equation}
where the volume averaged recombination time $t_{\rm rec}$ is:
\begin{eqnarray}
t_{\rm rec}&=&\!\left[C_{{\rm H_{II}}}\alpha_B(T_0)(1+Y/4X)\langle n_H\rangle(1+z)^3\right]^{-1}\nonumber\\
&\approx&\!0.93\,{\rm Gyr}\!\left(\frac{C_{{\rm H_{II}}}}{3}\right)^{-1}\!\!\!\left(\frac{T_0}{2\times10^4~{\rm K}}\right)^{0.7}\!\!\left(\frac{1+z}{7}\right)^{-3}\!\!\!\!\!\!\!,
\end{eqnarray}
where $\alpha_B(T_0)$ is the case B hydrogen recombination coefficient at $T_0=2\times10^4$~K, which takes the value $1.6\times10^{-13}~\rmn{cm}^3/\rmn{s}$, and $C_{\rm H_{II}}$ is the effective clumping factor in ionised gas in the diffuse IGM. There is some uncertainty in the value of $C_{\rm H_{II}}$ (see e.g. Fig. 5 of \citealt{Gnedin16}). Here we have used a constant value of 3, but note that we also tested the redshift dependent parametrization of \citep{Pawlik09}, and find that it only effects our final value of $\tau$ at the 3~per~cent level. Finally, $\dot{n}_{\rm ion}\equiv \rmn{d}n_{\rm ion}/\rmn{d}t$ is the globally averaged
rate of production of hydrogen ionizing photons:
\begin{equation}
\dot{n}_{\rm ion}=f_{\rm esc}\,\xi_{\rm ion}\,\rho_{\rm sfr},
\end{equation}
where $f_{\rm esc}$ is an effective fraction of photons produced by 
the stellar population that escape to ionise the IGM, $\xi_{\rm ion}$ is the ionizing
photon production efficiency per unit time per unit SFR for a typical stellar
population and takes the value ${\rm log}\xi_{\rm ion}=53.14$ where $\xi_{\rm ion}$ is measured in units of photons s$^{-1}/({\rm M}_\odot {\rm yr}^{-1})$ \citep{Robertson15}. Note that the relation between $\dot{n}_{\rm ion}$ and $\rho_{\rm sfr}$ has a degeneracy between $\xi_{\rm ion}$ and $f_{\rm esc}$. Therefore, although we focus below in a range of plausible values for $f_\rmn{esc}$, this range should be interpreted keeping in mind the degeneracy with $\xi_{\rm ion}$.

\begin{figure}
\includegraphics[width=0.49\textwidth]{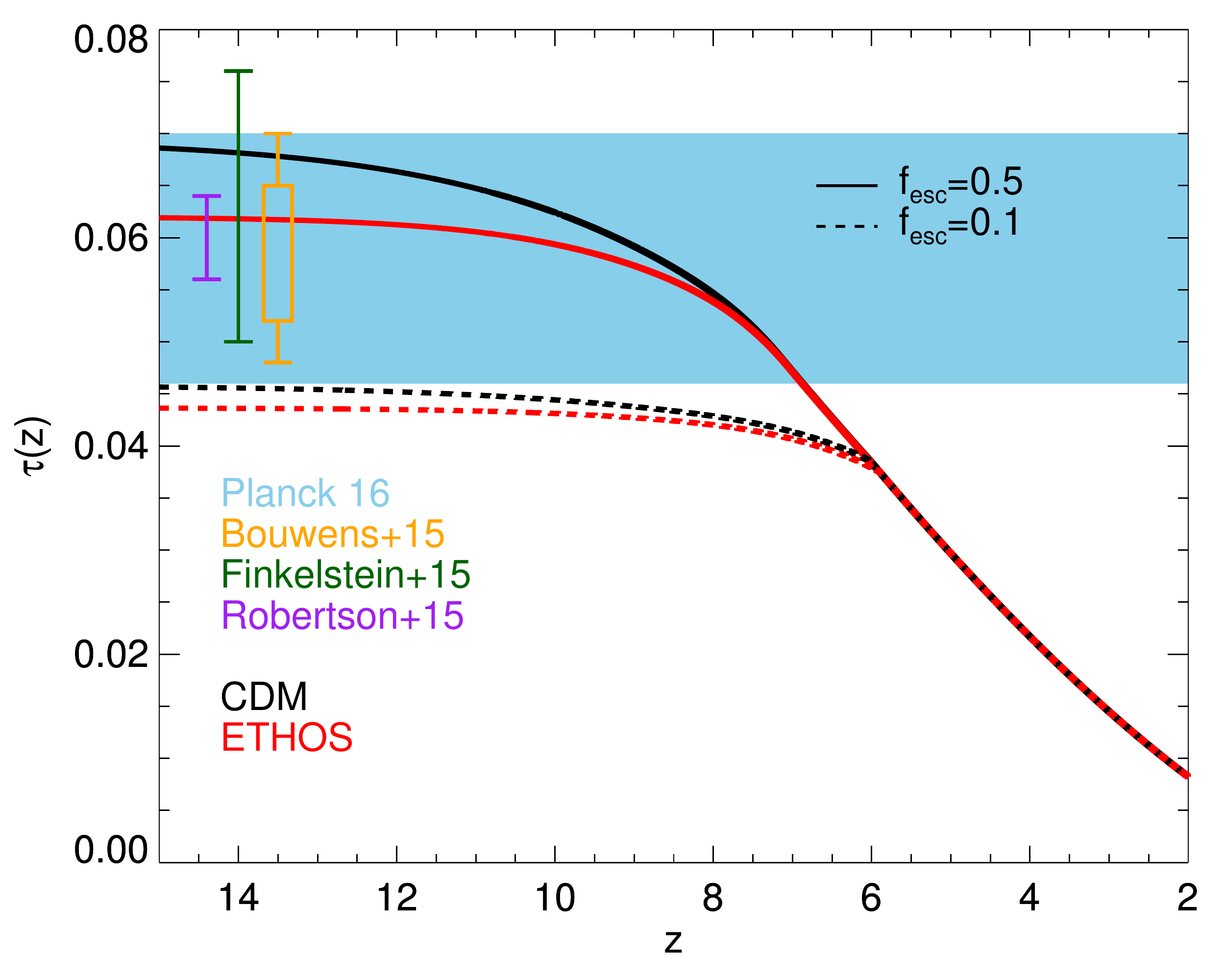}
\caption{Optical depth, $\tau(z)$ as a function of redshift. Black denotes CDM
and red ETHOS. We show calculations in which $f_\rmn{esc}=0.5$ (solid lines), and
$f_\rmn{esc}=0.1$ (dashed lines). The light blue region signifies
the allowed region measured in \citet{PlanckXLVI16}. The orange data point marks the 68~per~cent (box) and 95~per~cent (error bars) confidence regions from \citet{Bouwens15}.  The purple error bar shows the 68~per~cent confidence region measured by \citet{Robertson15}, while the green error bar is likewise the 68 per cent region of \citet{Finkelstein15}.}
\label{tauz}
\end{figure}

We take the SFR density directly from the simulations (with the extrapolation shown in Fig.~\ref{UVrhoExp}) as opposed to e.g. \cite{Robertson15}, who derived it from a maximum likelihood fit to
observations. 
To maximise the effect of reionization, we adopt a fraction of ionizing photons,
$f_\rmn{esc}$, that  can  escape  their  host  haloes  into  the  intergalactic
medium, equal to $f_\rmn{esc}=0.5$ (\citealp[as supported by e.g.][]{Fontanot14}, \citealp[ but see also][for some discussion of why a lower value may be preferred]{Wise14,Ma15}). In
order to calculate the escape fraction in a self-consistent way, we would need
to either perform simulations with radiative transfer
\citep[e.g.][]{Xu16,Gnedin16}, or post-process our snapshots using a hybrid
approach \citep{Ma15,Sharma16}. In general, the value of $f_\rmn{esc}$ varies
greatly temporally and across spatial regions. In the approach we are using,
the relevant quantity is an effective
redshift-dependent volume-average value of $f_\rmn{esc}$. For CDM,
\citet{Gnedin16} computes this value showing that it has a complex behaviour
with redshift with a value $\sim0.2$ at $z=7-9$, and a scatter of a factor of a
few depending on the clumping factor of the ionised gas. This value is
sensitive to the details of the baryonic physics implementation, and more
importantly, it would be different for ETHOS. For the purpose of this paper, we
choose a constant value $f_\rmn{esc}$, noting that is a relevant source of
uncertainty in computing the optical depth.
We present the resulting optical depth and its redshift behaviour in
Fig.~\ref{tauz}. 
    
In our maximal model, the value of $\tau(z=15)$ measured for ETHOS is only
$8$~per~cent lower than that of CDM. Both models are in good agreement with the constraints derived by Planck. We also compare our results
to the estimate of $\tau$ calculated from the high-redshift luminosity function
by \citet{Bouwens15}, \citet{Robertson15} and \citet{Finkelstein15}. Both CDM and ETHOS are consistent with these observations.

Setting the escape fraction to $0.1$ (dashed lines) reduces the value of $\tau$ for CDM
(ETHOS) by $31$~per~cent ($43$~per~cent). We also consider a minimal scenario in which ETHOS achieves the lower limit of the Planck measurement without any extrapolation in the SFR density, and find that we require $f_\rmn{esc}\ge0.14$ (not plotted). Overall, we conclude that both CDM and ETHOS are essentially consistent with
constraints on the optical depth across values of $f_{\rm esc}$ from $\sim$0.1 to 0.5, with the latter preferring slightly higher $f_{\rm esc}$.
We also note that the value of $f_\rmn{esc}$ is likely to be different between
the two models. For instance, it is suggested in \citet{Dayal17} that a
steeper redshift evolution of the ionizing photon escape fraction in WDM models
is a way to compensate for the cut-off in the power spectrum. The recombination
rate may also be different due to absorption from minihaloes ($10^{4}$-$10^{7}\msun$), which are
present in CDM but erased in ETHOS \citep{Yue12,Rudakovskyi16}. There will be additional uncertainty on our
results given the systematic uncertainties on the
baryon physics sub-resolution model, as also argued by \citet{VillanuevaDomingo18}.

Keeping all these caveats/uncertainties in mind, our results indicate that the ETHOS benchmark model, which was 
calibrated to alleviate the CDM small-scale challenges using dark-matter-only simulations in \citet{Vogelsberger16}, is consistent with 
high-redshift observables under reasonable assumptions about baryonic physics
in this mass regime.

\section{Conclusions}
\label{conc}
   
The particle properties of dark matter remain a mystery. 
Hidden dark matter particle interactions are motivated by a plethora of particle physics models where the dark sector possesses a richer phenomenology with several dark matter species and new forces. A promising search for such interactions lies in looking
for their dynamical signature in the formation and evolution of galaxies.
Particle models with hidden interactions have an astrophysical impact if they can either
(i) alter the primordial linear power spectrum (e.g., through a Silk-like
damping caused by dark matter interaction with relativistic particles in the
early Universe; e.g. \citealt{Boehm02,Buckley14,Boehm14}), or (ii) modify the
dark matter phase space density in the centre of galactic-size haloes (e.g.,
through strong dark matter self-interactions; e.g.
\citealt{Spergel00,Vogelsberger12,Rocha13,Zavala13}). These possibilities are
central to a recently proposed framework that generalises the theory of
structure formation by self-consistently mapping the parameters of allowed
particle physics models into effective parameters for structure formation
(ETHOS, \citealt{CyrRacine16,Vogelsberger16}). The ETHOS framework is a
powerful way to explore the consequences of new dark matter physics for galaxy
formation/evolution. For instance in \citet{Vogelsberger16}, dark matter-only simulations were used
to find a benchmark model (ETHOS-4, which we refer to as ETHOS in this paper for simplicity) that eases the tensions with some of the
outstanding small-scale challenges facing the standard CDM model in regards to
the properties of Milky Way satellites, i.e.,  their abundance and inferred
inner dark matter structure \citep[for a review see][]{bullock2017}.
   
In order to further explore and constrain the ETHOS framework, we have studied
here the consequences of this specific ETHOS benchmark model in the high redshift 
Universe, which is an an environment very different from that of the Milky Way. Our goal was to understand the consistency of the model with high-redshift observations. To
accomplish this, we have performed cosmological hydrodynamical simulations
with a well-developed galaxy formation model. 
The simulations cover a volume of $(36.2\Mpc)^3$, and each run employs a simulation dark matter particle mass of $1.76\times10^{6}\msun$ and dark matter softening length of
$724\,{\rm pc}$. The average gas cell mass is $2.69 \times 10^5\msun$ and the
gas softening length is adaptive with a minimum of $181\,{\rm pc}$. This mass resolution is comparable to the highest numerical resolution of any resolved uniform volume hydrodynamical simulations of an alternative dark matter model.

At high redshifts ($z\geq6$), the main differences between the benchmark ETHOS model and
the standard CDM model are caused by the former having a primordial cut-off in
the power spectrum due to dark matter-dark radiation interactions, suppressed
for wavenumbers $\gsim$14.5~$\rmn{Mpc}^{-1}$, with an oscillating amplitude at
higher wavenumbers (see Fig.~\ref{MPS}). This cut-off reduces the number density
of low mass haloes and also delays the onset of structure formation across
the mass hierarchy. These two phenomena lead to a suppression of galaxy
formation at low masses, and thus to a reduction of the available ionizing
photons responsible of re-ionizing the Universe. Both of these are connected to the observed
low-luminosity end of the UV luminosity function
\citep[e.g.][]{Bouwens15b,Livermore17,Ishigaki18}, and the optical depth of cosmic
microwave background photons, denoted by $\tau$ \citep{PlanckXLVI16}.  
   
We find that although the number of low mass galaxies is suppressed in
ETHOS relative to CDM, the difference is still indistinguishable in
current observations: $\sim0.1\,{\rm dex}$ for $M_U\sim-14.5$ at $z\sim8$ in the FUV luminosity function, while the observational errors ($1\sigma$) at similar redshifts and slightly brighter magnitudes are $\sim1\,{\rm dex}$ based on
\citealt{Livermore17,Ishigaki18} (see upper panel of Fig.~\ref{JWST_LF}). This leaves the prospects of progressively differentiating these models in this way to upcoming galaxy surveys, beginning with those planned for the JWST.  Based on our simulations, we have presented predictions for the rest-frame FUV (1500 nm) and NIR (1.15$\mu$m) luminosity functions, as well as in the observer-frame for one of the filters (F150W) of the NIRCam instrument in JWST (Fig.~\ref{JWST_LF}). Predictions for other filters are available upon request to the authors.

On the other hand, we also find that, for the mass range affected by the primordial cut-off of the power spectrum, high-redshift galaxies in ETHOS are
brighter per unit halo mass than is the case for CDM (see Fig.~\ref{MhUV23}), a result that is
consistent overall with recent studies based on a WDM cosmology coupled with 
semi-analytic model of galaxy formation \citep{Bose17a,Bose16c}. Since these results are based on very different models of galaxy formation and evolution, it suggests that having high-redshift low-mass galaxies with a higher efficiency of star formation is a generic feature of models with a cut-off in the primordial power spectrum.  

The brighter starbursts in ETHOS partially compensate for the deficit of UV photons
due to the low galaxy number density. This compensating effect reduces the
naive expectations of the impact of the cut-off in the power spectrum in the
optical depth $\tau(z)$. To estimate the optical depth from our simulations, we use the SFR measured directly from the gas properties in our simulations to compute the cumulative SFR density down to halo masses
of $10^8\msun$, which is our effective resolution limit in the halo mass function. By this halo mass, the SFR density has essentially converged to a maximum value which we use to estimate analytically the number density of ionizing photons
and thus the optical depth. We find that the bright ETHOS starburst
galaxies provide a boost to the optical depth at all $z>6$ over the naive expectations. Ultimately however, the
great number density of small galaxies in CDM wins out over the relative
brightness of their ETHOS counterparts, such that the total ETHOS optical depth
is still suppressed relative to CDM, but only by $\lesssim$10~per~cent within a range of values of the escape fraction in between 0.1 and 0.5 (see Fig.~\ref{tauz}). This suppression is relatively small compared to the uncertainties in both the experimentally measured optical depth of the CMB, the UV-photon escape fraction, and the ionizing photon production rate efficiency. Within the
assumptions of our method, we find that the CDM and ETHOS models are equally consistent with current observations. A significant improvement upon our results can only be achieved by a self-consistent calculation of the escape fraction in the ETHOS and CDM simulations; i.e., including radiative transfer that accounts for clumping and self-shielding, and a better understanding of the
impact of galaxy formation modelling in the faint end of the luminosity function at
very high redshifts.
   
We conclude that the ETHOS benchmark model, chosen to alleviate the small-scale
issues of CDM at the scale of satellite galaxies, is currently consistent with the
high-redshift abundance of galaxies, and with reionization constraints.
Limitations of this study include our baryonic mass resolution ($\sim 2\times 10^5\msun$ average mass cell), 
which is still too coarse to resolve the $U$-band luminosity function down
to the faintest galaxies responsible from reionization (particularly at $z>8$). This
adds uncertainty in our calculation of the global SFR density, which in this work is based on the SFR calculated in resolved haloes based on their gas content, but that lack the resolution to form the galaxies within. We have also not explored variations over the particular baryonic
physics implementation we have used. Studying the synergy between variations of
the dark and baryonic physics (i.e. varying the effective parameters in the
dark matter and baryonic physics sectors that impact galaxy formation and
evolution) is one of our near future plans. Finally, 
self-consistent radiation-hydrodynamics simulations (Kannan et al. in prep) are needed to explore in more detail the reionization history
and to make detailed predictions of the high-redshift galaxy population for CDM and non-CDM models.
   
\section*{Acknowledgements}

We thank Torsten Bringmann, Rachael Livermore, Steve Finkelstein and Christina Williams for useful comments and suggestions. We further thank Volker Springel for giving us access to {\sc Arepo}, and Jose O\~norbe for help with computing the optical depth. MRL is  supported  by  a  COFUND/Durham  Junior Research Fellowship under EU grant 609412. MRL and JZ acknowledge
support by a Grant of Excellence from the Icelandic Research Fund (grant number
173929$-$051). MV acknowledges support through an MIT RSC award, the support of the Alfred P.
Sloan Foundation, and support by NASA ATP grant NNX17AG29G. MBK acknowledges support from NSF grant AST-1517226 and from NASA grants NNX17AG29G and HST-AR-13888, HST-AR-13896, HST-AR-14282, and HST-AR-14554 from the Space Telescope Science Institute, which is operated by AURA, Inc., under NASA contract NAS5-26555. F.-Y. C.-R. acknowledges the support of the National Aeronautical and Space Administration ATP grant NNX16AI12G at Harvard University. Some numerical calculations were run on using allocation TG-AST140080 granted by the Extreme Science and Engineering Discovery Environment (XSEDE), which is supported by the NSF.

\bibliographystyle{mnras}


\bsp
\label{lastpage}

\end{document}